\documentclass[journal]{IEEEtran}
\usepackage{graphicx} 
\usepackage{multirow}
\usepackage{subfigure}
\usepackage{color,soul}
\usepackage{hhline}
\usepackage{xcolor}
\usepackage{transparent}
\usepackage{url}
\usepackage[noadjust]{cite}
\usepackage{booktabs}
\begin{document}

\title{Deep Learning-based Non-Intrusive Multi-Objective
Speech Assessment Model with Cross-Domain
Features}

\author{Ryandhimas E. Zezario,~\IEEEmembership{Student Member,~IEEE, } Szu-Wei Fu, Fei Chen,~\IEEEmembership{Senior Member,~IEEE}
\par
Chiou-Shann Fuh, Hsin-Min Wang,~\IEEEmembership{Senior Member,~IEEE, }and Yu Tsao,~\IEEEmembership{Senior Member,~IEEE }
 
\thanks{Ryandhimas E. Zezario is with the Department of Computer Science and Information Engineering, National Taiwan University, Taipei, Taiwan, and also with the Research Center for Information Technology Innovation, Academia Sinica, Taipei, Taiwan.}
\thanks{Szu-Wei Fu is with Microsoft, Vancouver, Canada}
\thanks{Fei Chen is with the Department of Electrical and Electronic Engineering, Southern University of Science and Technology of China, Shenzhen, China.}
\thanks{Chiou-Shann Fuh is with the Department of Computer Science and Information Engineering, National Taiwan University, Taipei, Taiwan.}
\thanks{Hsin-Min Wang is with the Institute of Information Science, Academia Sinica, Taipei, Taiwan.}
\thanks{Yu Tsao are with the Research Center for Information Technology Innovation, Academia Sinica, Taipei, Taiwan, corresponding e-mail: (yu.tsao@sinica.edu.tw). }
\thanks{}}

\markboth{}%
{Shell \MakeLowercase{\textit{et al.}}: Bare Demo of IEEEtran.cls for IEEE Journals}

\maketitle

\begin{abstract}

In this study, we propose a cross-domain multi-objective speech assessment model called MOSA-Net, which can estimate multiple speech assessment metrics simultaneously. More specifically, MOSA-Net is designed to estimate the speech quality, intelligibility, and distortion assessment scores of an input test speech signal. It comprises a convolutional neural network and bidirectional long short-term memory (CRNN) architecture for representation extraction, and a multiplicative attention layer and a fully connected layer for each assessment metric. In addition, cross-domain features (spectral and time-domain features) and latent representations from self-supervised learned (SSL) models are used as inputs to combine rich acoustic information from different speech representations to obtain more accurate assessments. Experimental results show that MOSA-Net can improve the linear correlation coefficient (LCC) by 0.026 (0.990 vs 0.964 in seen noise environments) and 0.012 (0.969 vs 0.957 in unseen noise environments) in perceptual evaluation of speech quality (PESQ) prediction, compared to Quality-Net, an existing single-task model for PESQ prediction, and improve LCC by 0.021 (0.985 vs 0.964 in seen noise environments) and 0.047 (0.836 vs 0.789 in unseen noise environments) in short-time objective intelligibility (STOI) prediction, compared to STOI-Net (based on CRNN), an existing single-task model for STOI prediction. Moreover, MOSA-Net, originally trained to assess objective scores, can be used as a pre-trained model to be effectively adapted to an assessment model for predicting subjective quality and intelligibility scores with a limited amount of training data. Experimental results show that MOSA-Net can improve LCC by 0.018 (0.805 vs 0.787) in mean opinion score (MOS) prediction, compared to MOS-SSL, a strong single-task model for MOS prediction. In light of the confirmed prediction capability, we further adopt the latent representations of MOSA-Net to guide the speech enhancement (SE) process and derive a quality-intelligibility (QI)-aware SE (QIA-SE) approach accordingly. Experimental results show that QIA-SE provides superior enhancement performance compared with the baseline SE system in terms of objective evaluation metrics and qualitative evaluation test. For example, QIA-SE can improve PESQ by 0.301 (2.953 vs 2.652 in seen noise environments) and 0.18 (2.658 vs 2.478 in unseen noise environments) over a CNN-based baseline SE model.
\end{abstract}


%
\IEEEpeerreviewmaketitle

\section{Introduction}
\IEEEPARstart
SPEECH assessment metrics are indicators that quantitatively measure the specific attributes of speech signals. These metrics are vital to the development of speech-related application systems. A direct assessment approach measures the difference between the distorted/processed speech and clean reference at the signal level. The speech distortion index (SDI) \cite{sdi} is a well-known example that calculates the distortion of the distorted/processed speech compared with the clean speech. Meanwhile, the signal-to-noise-ratio (SNR) \cite{snr} and segmental SNR (SSNR) \cite{ssnr} are other well-known metrics that indicate the difference between processed and noisy speech. Scale-invariant source-to-noise ratio (SI-SNR) \cite{sisnr} and optimal scale-invariant signal-noise ratio (OSI-SNR) \cite{osisnr} are improved versions of SNR that have also been proven effective in assessing speech signals. Although these signal-level metrics can directly indicate the distortion or SNR of the distorted/processed speech compared to the clean reference, they may not fully reflect the quality and intelligibility of the distorted/processed speech. Therefore, many evaluation metrics have been proposed for measuring speech quality and intelligibility. 
\par
Existing speech quality and intelligibility evaluation metrics can be classified into two categories: subjective and objective metrics. The subjective evaluation metrics are based on test scores from human listeners. To obtain subjective scores, speech samples are played to a group of human subjects, and these subjects provide feedback regarding the quality or intelligibility levels of the played speech signals. In terms of speech quality, the mean opinion score (MOS) is a typical numerical indicator in listening tests. In most cases, the MOS metric categorizes speech quality into five levels, ranging from one to five, with a higher score indicating better quality. By contrast, the intelligibility score refers to the ratio of the number of accurately recognized words/phonemes/sentences in the played speech samples, where in this study, we focus on word-level speech intelligibility scores. To attain an unbiased assessment of speech quality and intelligibility, it is necessary to recruit a sufficient number of human subjects, and each subject must listen to a significant amount of speech utterances encompassing diverse acoustic conditions, including speakers and distortion sources. This testing strategy is prohibitive and may not always be feasible. Hence, several objective evaluations metrics have been developed as surrogates for human listening tests \cite{ref_19,ref_20, ref_22,ref_25,ref_26,ref_32,ref_33,ref_34,ref_21, ref_28,ref_29,ref_30,ref_31,ref_23, ref_24,ref_27,ref_35,ref_36,ref_37,ref_38,ncm,csii,ref_39,estoi,moda,srmr}.
\par

%

Generally, a conventional objective quality evaluation metric comprises two stages. The first stage includes a series of signal processing units designed to convert speech waveforms into handcrafted acoustic/auditory features. The second stage derives a mapping function to predict the speech quality score based on acoustic/auditory features. The mapping function can be implemented via linear regression \cite{ref_22}, polynomial regression \cite{ref_19, ref_20}, multivariate adaptive regression spline \cite{ref_21}, and machine learning methods, such as Gaussian mixture models \cite{ref_25, ref_30,ref_31}, support vector regression \cite{ref_26, ref_28}, and artificial neural networks \cite{ref_23, ref_24, ref_27}. Depending on whether clean reference speech is required, objective speech quality metrics can be further classified into two categories: intrusive metrics \cite{ref_19} and non-intrusive metrics \cite{ref_20, ref_22,ref_25,ref_26,ref_32,ref_33,ref_34,ref_21, ref_28}. Well-known intrusive metrics include perceptual evaluation of speech quality (PESQ) \cite{ref_19} and perceptual objective listening quality analysis (POLQA) \cite{polqa_2013}. Compared with intrusive metrics, non-intrusive metrics do not require a clean reference; therefore, they are more suitable for real-world scenarios, but generally have lower assessment capabilities.  
\par
Objective intelligibility evaluation metrics can be classified into two categories. One category first segregates the speech signal under analysis into frequency subbands, and assumes that each subband independently contributes to the intelligibility. Next, the long-term subband SNR is calculated and then normalized to a value between 0 and 1. Finally, the intelligibility score is obtained using the perceptually weighted average of the normalized subband SNRs. Notable examples of this category include the articulation index (AI) \cite{ref_35}, speech intelligibility index (SII) \cite{ref_36}, extended SII (ESII) \cite{ref_37}, and coherence SII (CSII) \cite{csii}. The other category is derived based on the observation that reverberation and/or additive noise tends to reduce the modulation depth of the distorted signal, compared with the clean reference signal. Well-known approaches of this category include the speech transmission index (STI) \cite{ref_38}, spectro-temporal modulation index (STMI), normalized-covariance measure (NCM) \cite{ncm}, short-time objective intelligibility (STOI) \cite{ref_39}, extended STOI (eSTOI) \cite{estoi}, spectrogram orthogonal polynomial measure (SOPM) \cite{somr}, neurogram orthogonal polynomial measure (NOPM)\cite{nopm}, neurogram similarity index measure (NSIM) \cite{nsim} and weighted spectro-temporal modulation index (wSTMI) \cite{edraki2020speech}. To avoid the necessity for clean reference speech, several non-intrusive approaches have been proposed. Most of them adopt statistical models of clean speech signals or psychoacoustic features for speech understanding \cite{rev_int}. Notable non-intrusive speech intelligibility metrics include modulation-spectrum area (ModA) \cite{moda}, speech-to-reverberation modulation energy ratio (SRMR) \cite{srmr}, and the non-intrusive STOI \cite{ref_43}. 
\par
Recently, the emergence of deep learning algorithms has resulted in the development of many deep learning-based speech assessment models. These models are trained to predict human subjective ratings \cite{ref_44,ref_45,ref_46,ref_47,ref_48, dnsmos} or objective evaluation scores, in terms of speech quality \cite{ref_49, ref_55, ref_56} and intelligibility \cite{ref_52, ref_56}. To attain a higher assessment accuracy, the MBNet adopts the BiasNet architecture to compensate for the biased scores of a certain judge \cite{mbnet_mos}, In addition, the multi-task learning criterion that simultaneously optimizes multiple metrics is used to train the assessment model \cite{ref_56, choi2020neural}. Meanwhile, different acoustic features are used as input to the assessment model to consider information from different acoustic domains \cite{hu2021svsnet, wav2vec_mos}. 
\par
In contrast to current deep learning-based speech assessment models that use a single acoustic feature type and one particular objective to build the model, the proposed MOSA-Net aims to exploit rich acoustic information from multiple domains and utilize multiple objectives for model training. More specifically, MOSA-Net uses three different types of features: traditional spectral features, waveforms processed by learnable filters (based on the Sinc convolutional network \cite{sincnet}), and latent representations from self-supervised learned (SSL) models (wav2vec 2.0 \cite{wav2vec} and HuBERT \cite{hubert}). Note that wav2vec2.0 learns context information by identifying representations to the true quantized latent speech representation \cite{wav2vec}, and HuBERT predicts (hidden) cluster assignments \cite{hubert}. Instead of directly using the outputs, we use the embeddings of these SSL models as the SSL features. For more details, please refer to \cite{wav2vec_mos} and \cite{ssl-mos}. Additionally, MOSA-Net adopts a multi-task learning criterion that simultaneously predicts multiple objective assessment metrics, including speech quality, intelligibility, and distortion scores. MOSA-Net is composed of a convolutional neural network (CNN) and a bidirectional long short-term memory (BLSTM) with an attention mechanism. We systematically evaluated the performance of MOSA-Net based on various model architectures, training targets, acoustic features, and datasets. Experimental results (in terms of mean square  error (MSE), linear correlation coefficient (LCC), and Spearman’s rank correlation coefficient (SRCC)) demonstrate the advantages of cross-domain features, multi-tasking learning, and attention mechanism. Furthermore, experimental results also show that MOSA-Net, originally trained to predict objective speech assessment scores (i.e., PESQ, STOI, and SDI), can serve as a pre-trained model to be adapted to an assessment model with limited training data to predict subjective quality and intelligibility scores. To the best of our knowledge, this is the first work that adapts a pre-trained objective metric assessment model to a new model that can predict human subjective ratings with a small amount of adaptation data. This approach is different from most existing works that perform pseudo-labeling on the same metrics/tasks, such as DNSMOS \cite{dnsmos}. 
\par
In the literature, there have been several studies incorporating speech assessment models to improve SE performance \cite{fu2019metricGAN, metricgan+, DeepNoise, metricganu}, such as MetricGAN \cite{fu2019metricGAN} and MetricGAN+ \cite{metricgan+}. In addition, some SE methods prepare multiple SE systems and use speech assessment models to select the SE system that is most suitable for the test utterance, such as SSEMS \cite{zezario2019specialized} and ZMOS \cite{ref_54}. In contrast to existing approaches, we propose integrating the latent representation of MOSA-Net into the SE system, and derive a novel quality-intelligibility-aware (QIA)-SE system. Experimental results show that QIA-SE achieves notable improvements over the baseline SE systems and several existing SE systems, which confirms the advantage of combining the knowledge in the speech assessment model to improve the enhancement capability.
\par
The remainder of this paper is organized as follows. We first review related work in Section \ref{sec:related}. Subsequently, we elaborate the proposed methods in Section \ref{sec:methodology}. In Section \ref{sec:experiments}, we describe the experimental setup, report the experimental results, and discuss our findings. Finally, we conclude our work in Section \ref{sec:conclusion}.

\section{RELATED WORK}
\label{sec:related}

\subsection{Deep Learning-based Assessment Metrics}
To date, deep learning models have been widely used to build speech assessment systems. In this section, we review several deep learning-based assessment metrics based on different targets and model architectures. 

As mentioned earlier, the assessment targets can be classified into two types. The first type is human subjective ratings, and the second is objective assessment scores. When the target is the human subjective ratings, the learned assessment metric through appropriate modeling can directly predict the human subjective ratings \cite{mbnet_mos, wav2vec_mos, ref_51, dnsmos, andersen2018nonintrusive, SIP_DL}. However, a significant number of subjective listening tests encompassing many listeners and acoustic conditions must be conducted in advance to prepare ground-truth labels for an unbiased training set. In addition, it is difficult to extend the dataset to new domains, because additional subjective tests must be conducted. According to the training target criterion, the human subjective ratings can be classified into two categories: quality \cite{ref_51, dnsmos, mbnet_mos} and intelligibility scores  \cite{MONO99, somr, dubno2005word}. Notable systems associated with subjective quality metrics include the following: (1) The MOSNet \cite{ref_51}, which combines the utterance-level and frame-level scores to estimate the MOS of an utterance; (2) the DNSMOS \cite{dnsmos}, which uses the teacher-student architecture to eliminate subjective bias; and (3) the MBNet \cite{mbnet_mos}, which compensates for individual judgement biases using a BiasNet architecture. Compared with speech quality assessment, there is less work on predicting subjective intelligibility scores using deep learning models. For example, (1) Andersen et al. \cite{andersen2018nonintrusive} used a CNN model to accommodate the entire signal composed of multiple sentences to estimate the scalar value of the intelligibility score, and (2) Pedersen et al. \cite{SIP_DL} used a CNN architecture to calculate scores locally in a short time to achieve more efficient learning with limited listening test data. 
\par
The second group adopts objective speech assessment metrics as the ground-truth labels for model training. Similar to the first group, the objective speech assessment metrics can also be divided into two categories: quality and intelligibility. For objective speech quality assessment, the PESQ \cite{ref_49, ref_55, ref_56}, POLQA \cite{POLQA}, and HASQI \cite{ref_56} scores obtained by comparing the test speech with the reference speech are often used as the ground-truth scores for training the deep-learning-based assessment metrics. For speech intelligibility assessment, the STOI \cite{ref_52, ref_56, ref_55} and hearing-aid speech perception index (HASPI) \cite{katehaspi} scores are used as the training targets.  
\par
Many model architectures have been used to construct the deep learning-based assessment metrics, e.g., BLSTM \cite{ref_49}, pyramid BLSTM \cite{pyramidlstm}, CNN \cite{ref_47, dnsmos}, and CNN-BLSTM \cite{ref_51, mbnet_mos}. In addition, attention mechanism \cite{ref_52, ref_56 }, multi-task learning \cite{ref_56, choi2020neural}, and additional network that compensate for score biases \cite{mbnet_mos} have been used to improve assessment capabilities. In terms of input, different acoustic features have been explored, which can be classified into three categories. The first category includes traditional spectral features such as log Mel features \cite{dnsmos} and power spectral (PS) features \cite{ref_49,ref_51,ref_52}. The second category uses learnable filters to extract features from raw waveform \cite{ref_55,end2endmultisubject}. The third category is based on the end-to-end features of the self-supervised learned network \cite{wav2vec_mos}.

\par

\subsection{Incorporating Speech Assessment Metrics to SE}
The idea of incorporating informative latent representations from pre-trained models to guide target speech processing tasks has been extensively studied. For example, SE systems using speaker embedding \cite{ref_58,ref_59,ref_60,ref_61} and noise embedding \cite{ref_62} have been shown to provide improved SE performance. Since the goal of speech assessment metrics is to estimate speech quality/intelligibility attributes given a distorted/processed speech signal, it is feasible to use the information from these assessment metrics to guide the SE process to achieve better speech quality and intelligibility. These approaches can be classified into two categories. The first category directly uses speech assessment metrics as training targets to train the SE system \cite{fu2019metricGAN, metricgan+, nayem21_interspeech}. The second category uses assessment metrics to determine the best model architecture or select the most appropriate output  \cite{chang2021moevc, ref_54, zezario2019specialized}. 

\graphicspath{ {./images/} }
\begin{figure}[t]
\centering
\includegraphics[width=8cm]{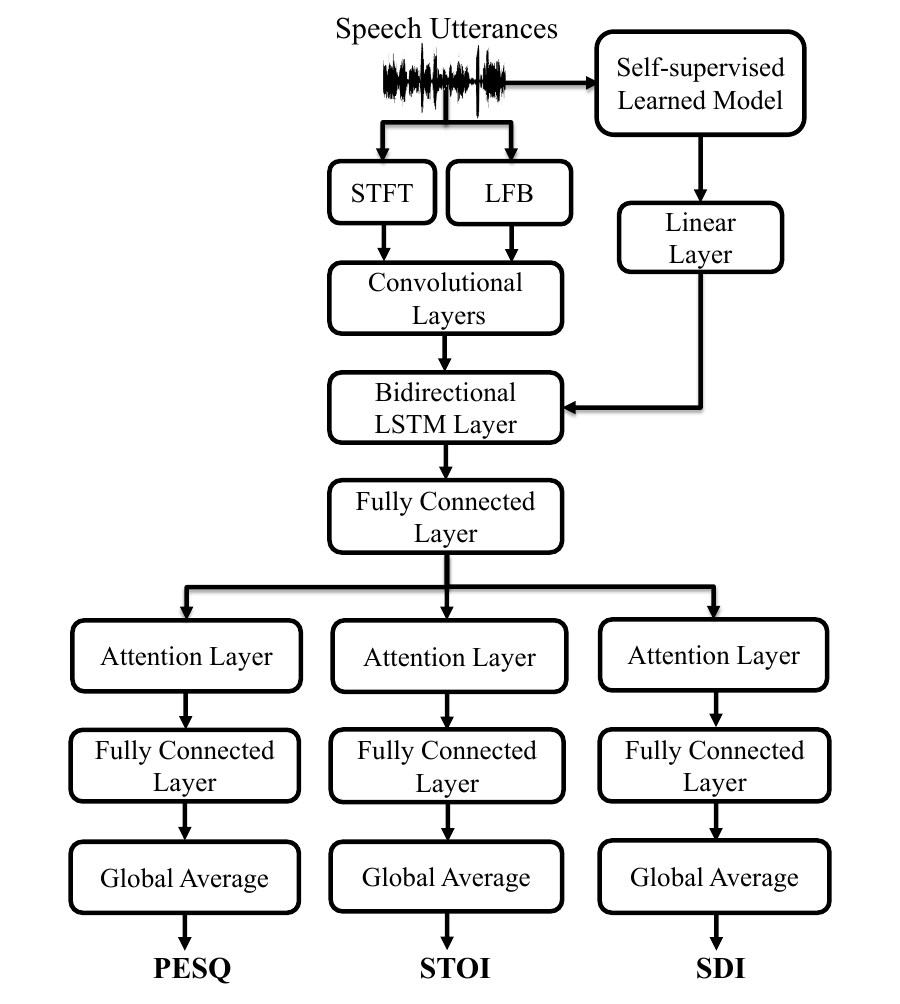} 
\caption{Architecture of the MOSA-Net model.} 
\label{fig:MOSA-Net}
\end{figure}
\section{PROPOSED METHODS}
\label{sec:methodology}
In this section, we first present the proposed MOSA-Net model. Subsequently, we will explain how to use latent representations for obtaining better speech quality or intelligibility.

\subsection{Multi-Target Speech Assessment Model with Cross-Domain Features (MOSA-Net)} 
Fig. 1 shows the overall architecture of the MOSA-Net model. As shown in the figure, MOSA-Net adopts cross-domain acoustic features and predicts multiple assessment scores. Given a speech waveform $\textbf{X}=[x_1,…,x_n…,x_N ]$,
the model takes two branches of the input. In the first branch, the speech waveform $\textbf{X}$ is processed by STFT and learnable filter banks (LFB) separately, and the estimated spectral and filtered signal features are fed into a convolutional layer. In the second branch, the speech waveform $\textbf{X}$ is processed by a self-supervised learned model (HuBERT \cite{hubert} or wav2vec 2.0 \cite{wav2vec}). It is noteworthy that different types of features are concatenated by temporal dimension. That is, the total number of frames of an utterance is the summation of the frame numbers of the PS, LFB, and SSL features. Besides, linear transformations are used to make these features have the same feature dimension. The combined features are further processed by a bidirectional layer and a fully connected layer. Subsequently, a set of attention layers is used for the corresponding objective assessment metrics. In our implementation, multiplicative attention is used in the attention layers because of its high efficiency and decent performance. Next, for each metric, a fully connected layer is used to generate the frame-wise scores. Finally, based on the frame-level scores, a global average operation is applied to calculate the final predicted PESQ, STOI, and SDI scores.

\par
Considering that speech utterances may contain stationary and/or non-stationary noise in different segments of frames, directly estimating the utterance level score may result in less accurate estimation. Therefore, the MOSA-Net aims to combine utterance-level and frame-level score estimations. Accordingly, the objective function of MOSA-Net is defined as follows:

\begin{equation}
\label{eq:loss}
   \small
    \begin{array}{c}
   L_{All} = \gamma_{1}L_{PESQ} + \gamma_{2}L_{STOI} + \gamma_{3}L_{SDI}
    \\
    L_{PESQ}=\frac{1}{N}\sum\limits_{n=1}^N [(Q_n-\hat{Q}_n)^2+\frac{\alpha_Q}{L(U_n)}\sum\limits_{l=1}^{L(U_n)}(Q_n-\hat{q}_{nl})^2]
    \\
    L_{STOI}=\frac{1}{N}\sum\limits_{n=1}^N [(I_n-\hat{I}_n)^2+\frac{\alpha_I}{L(U_n)}\sum\limits_{l=1}^{L(U_n)}(I_n-\hat{i}_{nl})^2]
    \\
    L_{SDI}=\frac{1}{N}\sum\limits_{n=1}^N [(S_n-\hat{S}_n)^2+\frac{\alpha_S}{L(U_n)}\sum\limits_{l=1}^{L(U_n)}(S_n-\hat{s}_{nl})^2]
    \end{array} 
\end{equation}
where ${Q_n}$, ${I_n}$, and ${S_n}$ are the true utterance-level scores of the PESQ, STOI, and SDI, respectively, of the $n$-th training utterance; $\hat{Q}_n$, $\hat{I}_n$, and $\hat{S}_n$ are the predicted utterance-level scores of the PESQ, STOI, and SDI, respectively, of the $n$-th training utterance; $N$ denotes the total number of training utterances; ${L(U_n)}=L(X_n)+{L(F_n)}+{L(C_n)}$ denotes the total number of frames in the $n$-th training utterance; $L(X_n)$, ${L(F_n)}$, and ${L(C_n)}$ are the number of frames of the STFT, LFB, and SSL features, respectively; $\hat{q}_{nl}$, $\hat{i}_{nl}$, and $\hat{s}_{nl}$ are the predicted frame-level PESQ, STOI, and SDI scores of the $l$-th frame of the $n$-th training utterance, respectively; as shown in Eq. (\ref{eq:loss}), for the $n$-th training utterance, there are ${L(U_n)}$ predicted frame-level PESQ, STOI, and SDI scores; $\alpha_Q$, $\alpha_I$, and $\alpha_S$ determine the weights between utterance-level and frame-level losses; and $\gamma_{1}$, $\gamma_{2}$, and $\gamma_{3}$ determine the weights between PESQ, STOI, and SDI losses. In Eq. \ref{eq:loss}, for each metric, the first and second terms estimate the accuracy of the utterance-level score and the frame-level score, respectively. The frame-level STOI, PESQ, and SDI ground-truth scores are the same as the utterance-level STOI, PESQ, and SDI ground-truth scores, respectively. Although $\gamma_{1}$, $\gamma_{2}$, $\gamma_{3}$, $\alpha_Q$, $\alpha_I$, and $\alpha_S$ can be adjusted, we set them all to 1 in our experiments. A linear layer was applied to reduce the high-dimensional SSL features to the same dimension as the STFT and LFB features.

\subsection{QIA-SE Model} 
\begin{figure}[t]
\centering
\includegraphics[width=5.5cm]{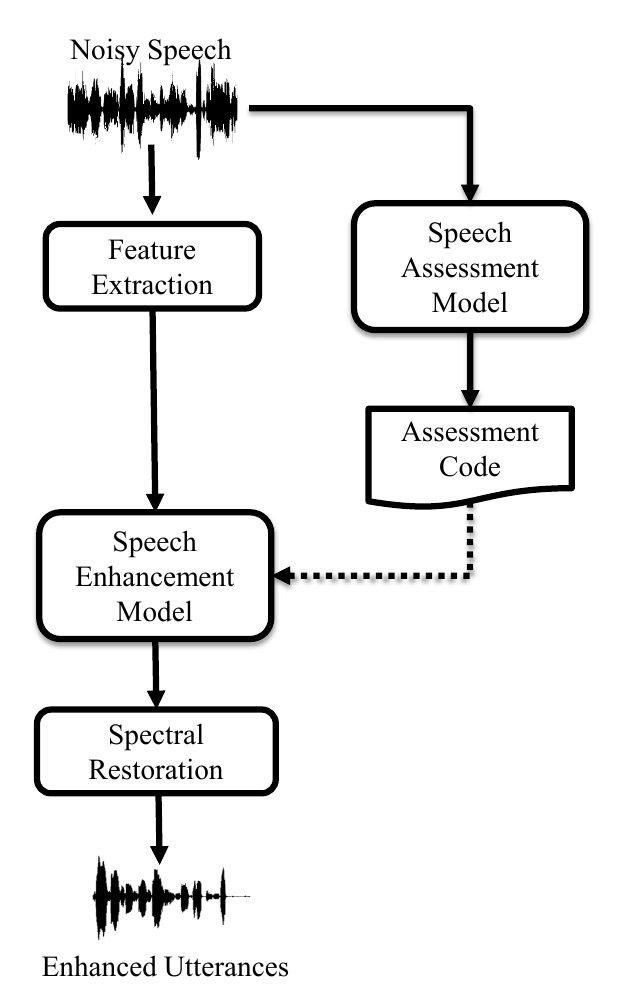} 
\caption{Architecture of the QIA-SE model.} 
\label{fig:QIA-SE}
\end{figure}

The QIA-SE model is designed to incorporate the latent representation from MOSA-Net to guide the SE process. The overall QIA-SE architecture is illustrated in Fig. \ref{fig:QIA-SE}. As shown in the figure, the noisy speech waveform is first converted to spectral features, $\textbf{Y}=[y_1,…,y_n…,y_L ]$, where $L$ is the total number of frames. QIA-SE aims to convert $\textbf{Y}$ to enhanced spectral features $\hat{\textbf{X}}$, referring to the latent representation features $\textbf{A}=[a_1,…,a_n…,a_L]$ extracted by the MOSA-Net from the input noisy speech waveform. 
In our current implementation, $\textbf{A}$ is the output of the fully-connected layer after attention in MOSA-Net, and is incorporated into the middle layer of QIA-SE via vector concatenation to guide the enhancement process,

\graphicspath{ {./images/} }

\begin{equation} \\
    \begin{array}{c}
    \textbf{H}_{1} \ = \ F_{\theta}^{(1)} \ (\textbf{Y}) 
    \\
    .
    .
    \\
    \textbf{H}_{k} \ = \ F_{\theta}^{(k)} \ (\textbf{H}_{k-1}) 
    \\
    \textbf{H}_{k+1} \ = \ F_{\theta}^{(k+1)} \ ([\textbf{H}^T_k,\textbf{A}^T]^T)
    \\
    .
    .
    \\
    \textbf{H}_{K} \ = \ F_{\theta}^{(K)} \ (\textbf{H}_{K-1})
    \\
    \hat{\mathbf{X}} \ = \ F_{\theta}^{(K+1)} \ (\textbf{H}_{K})
    \end{array} 
    \label{eq:QIA-SE}
\end{equation} 
where $k$ denotes the layer index, and $F_{\theta}^{(k)}$ indicates the transformation model of the $k$-th hidden layer. Parameter $\theta$ is optimized by minimizing the following loss function based on the MSE: 

\begin{equation}
\theta=\mathop{\arg\min}_{\theta}L(\mathbf{X}, \hat{\mathbf{X}}),
\end{equation}
where $\mathbf{X}$ denotes the clean speech reference. 
In this work, QIA-SE was built with a CNN model \cite{ref_54}, which comprised 12 convolutional layers, followed by a fully connected layer consisting of 128 neurons. Each convolutional layer contained four channels \{16, 32, 64, 128\} with three types of strides \{1, 1, 3\} in each channel.
Note that during training, MOSA-Net is frozen, and the parameters in the SE model are optimized. Since the input of the SE model is the STFT features, we only apply the MOSA-Net trained on the STFT features. We can take $\textbf{A}$ from the PESQ, STOI, or SDI branch, or from multiple branches by concatenating the representations by feature dimension. In the testing stage, noisy speech is first input into MOSA-Net to generate the latent representation $\tilde{\textbf{A}}$, and then input into the QIA-SE model to obtain the enhanced spectral features, as defined in Eq. \ref{eq:QIA-SE}. The enhanced speech waveform is generated by performing ISTFT on the enhanced spectral features along with phase information from the noisy speech.

\section{EXPERIMENTS}
\label{sec:experiments}
\subsection{Analysis of MOSA-Net}
In this section, we systematically investigate the correlations between the performance of MOSA-Net and different input features, model architectures, and output labels. A fair comparison of MOSA-Net with related neural evaluation metrics is presented.

\subsubsection{Experimental setup}
We used the Wall Street Journal (WSJ) dataset \cite{ref_63}, which comprises 37,416 training utterances and 330 test utterances. The training and test utterances were recorded at a sampling rate of 16 kHz. 
\par
We artificially contaminated the clean training utterances with 100 types of noises \cite{ref_64} at 31 different SNR levels, ranging from -10 to 20 dB with an interval of 1 dB, to prepare noisy utterances. We used 37,416 noisy-clean utterance pairs (randomly sampled a corresponding noisy utterance for each clean training utterance) to train an SE system, which was constructed by a BLSTM model with two bidirectional hidden layers, each containing 300 neurons. Then, we used the SE model to prepare enhanced utterances. Finally, we randomly sampled 1,500 clean utterances and corresponding 15,000 noisy utterances and 15,000 enhanced utterances and computed their PESQ, STOI, and SDI scores to form the training set for the MOSA-Net model.  
\par
We prepared two test sets: a seen test set and an unseen test set. For the seen test set, we randomly selected 300 clean utterances from the utterances other than the 1500 utterances in the training set and their corresponding 2,350 noisy utterances and 2,350 enhanced utterances. For the unseen test set, we selected 300 utterances from the test set of the WSJ dataset and artificially contaminated them with four unseen noise types (i.e., car, pink, street, and babble) at six SNR levels (i.e., -10, -5, 0, 5, 10, and 15 dB), amounting to 7,200 noisy utterances. The same SE model was applied to generate enhanced utterances. Note that the speakers in the unseen test set were not involved in the training set. We randomly selected 2,350 noisy utterances and 2,350 enhanced utterances together with the 300 clean utterances to form the unseen test set.
\par
To evaluate the proposed MOSA-Net model, we adopted three evaluation metrics, namely the MSE, LCC, and SRCC \cite{srcc}. Lower MSE scores indicate that the predicted scores are closer to the ground-truth assessment scores (the lower the better), whereas higher LCC and SRCC scores indicate that the predicted scores are of higher correlations to the ground-truth assessment scores (the higher the better).

\subsubsection{MOSA-Net with different model architectures}
First, we compared the MOSA-Net with different model architectures, including the CNN \cite{fu2019metricGAN}, BLSTM \cite{ref_49}, CNN-BLSTM \cite{ref_51}, and CNN-BLSTM with attention\cite{ref_52}. In the following, the MOSA-Net implemented with CNN-BLSTM will be denoted as CRNN, while the MOSA-Net implemented with CNN-BLSTM with attention will be denoted as CRNN+AT. In our implementation of the CNN, CRNN, and CRNN+AT models, we used a batch size of one and the Adam optimizer \cite{KingmaB14} with a learning rate of 0.0001. BLSTM was trained with the RMSprop optimizer \cite{tieleman2012lecture} with a learning rate of 0.001. For a fair comparison, we adopted the same acoustic features PS (power spectrogram) and a single-metric (either the PESQ or STOI score) learning criterion to train the model. To extract the PS features, each speech waveform was converted into a 257-dimensional spectrogram by applying a 512-point STFT with a Hamming window of 32 ms and a hop of 16 ms. The results of the MOSA-Net using the CNN, BLSTM, CRNN, and CRNN+AT are shown in Table I, where the results of both the seen and unseen tests are reported. For CNN, the model was constructed by convolutional layers completely. As shown in Fig. 1, the CRNN+AT model included 12 convolutional layers, each comprising four channels {16, 32, 64, and 128}, a one-layered BLSTM (with 128 nodes), and a fully connected layer (with 128 neurons). An attention layer was used to estimate the assigned objective assessment metric. Finally, the output of the attention layer was forwarded to a fully connected layer (with one neuron), and a global average operation was applied to generate the prediction score. The CRNN model architecture resembled CRNN+AT, where no attention layer was involved. For the CNN, we used the same model architecture as that reported in \cite{fu2019metricGAN}. The model comprised of four two-dimensional convolutional layers with the following filters and kernels configurations: [15, (5, 5)], [25, (7,
7)], [40, (9, 9)], and [50, (11, 11)]. In addition, the two-D global average pooling was added to fix feature dimension into 50, and the feature was mapped into three fully connected layers with the following configurations: 50 and 10 LeakyReLU nodes, and one linear node. For BLSTM, we used the same model architecture as that reported in \cite{ref_49}. The model comprised of one bidirectional LSTM layer with 100 nodes, followed by two fully connected layers with 50 exponential linear unit (ELU) nodes and one linear node.

\begin{table}[t]
\caption{LCC, SRCC, and MSE results of MOSA-Net using CNN, BLSTM, CNN-BLSTM (CRNN) and CRNN with attention (CRNN+AT) model architectures. The PS features are used as the input, and a single metric (either the PESQ or STOI score) is used to train MOSA-Net.}
\footnotesize
\begin{center}
 \begin{tabular}{c||c||c||c||c||c||c} 
 \hline
 \hline
  \multirow{2}{*}{\textbf{Model}} & \multicolumn{3}{c||}{Seen Noises} &  \multicolumn{3}{c} {Unseen Noises}  \\ \cline{2-7}
  
  &\textbf{LCC} & \textbf{SRCC} & \textbf{MSE}&\textbf{LCC} & \textbf{SRCC} & \textbf{MSE}  \\ [0.5ex] \cline{2-7}
 \hline\hline
  \multicolumn{7}{c} {PESQ score prediction} \\
 \hline
BLSTM&0.964&0.945&0.074&0.957&0.932&\textbf{0.075}\\ \hline
CNN&0.975&0.959&0.055&0.947&0.931&0.117\\ \hline
CRNN&0.981&0.965&0.042&\textbf{0.966}&0.949&0.078\\ \hline
CRNN+AT&\textbf{0.982}&\textbf{0.967}&\textbf{0.040}&0.965&\textbf{0.954}&0.092\\ \hline
 
 \hline
  \multicolumn{7}{c} {STOI score prediction} \\
 \hline
BLSTM&0.923&0.929&0.005&0.764&0.784&0.029\\ \hline 
CNN&0.936&0.939&0.004&0.698&0.694&\textbf{0.012}\\ \hline
CRNN&0.964&0.962&0.002&0.789&0.797&0.016\\ \hline
CRNN+AT&\textbf{0.970}&\textbf{0.968}&\textbf{0.001}&\textbf{0.827}&\textbf{0.815}&0.015\\ \hline

\end{tabular}
\end{center}
\end{table}
\par

As shown in Table I, CRNN slightly outperformed the CNN and BLSTM, in terms of both the PESQ and STOI predictions for the seen and unseen test sets. The results suggest that combining the abilities of CNN in extracting local invariant features and BLSTM in characterizing temporal characteristics can yield better performance than using individual CNN and BLSTM in this task. Additionally, CRNN+AT outperformed CRNN. This indicates that by incorporating the attention mechanism, the model can focus on the more important regions and hence allow MOSA-Net to achieve better prediction performance. 


Next, we compared the CRNN+AT model with two existing systems, Quality-Net \cite{ref_49} (based on BLSTM) and STOI-Net (based on CRNN) \cite{ref_52}. Note that the CRNN+AT model is actually a single-task MOSA-Net model, and the single-task MOSA-Net model with a single type of features (PS) is the same as the CRNN+AT based STOI-Net model in \cite{ref_52}.  
To qualitatively analyze the advantages of CRNN+AT, we used scatter plots \footnote{We drew the scatter plots using the open-source tool: https://seaborn.pydata.org/generated/seaborn.regplot.html. The regression lines were estimated with a confidence interval of 95\%.} to compare the prediction results of these models. Scatter plots can show the correlation between predicted and true scores. More specifically, if an assessment model can accurately predict the metric scores, the points will be densely distributed on the diagonal; otherwise, the points will be spread out from the diagonal.
As shown in Fig. 3, we can note that the predicted PESQ and STOI scores by MOSA-Net are more correlated with the true scores than Quality-Net and STOI-Net as the points of MOSA-Net are more densely distributed on the diagonal. More comparisons of MOSA-Net, Quality-Net, and STOI-Net are shown in Table II. From the table, we can again confirm that MOSA-Net outperforms these two previous works in almost all LCC, SRCC, and MSE metrics (except for the MSE metric for PESQ prediction under the unseen noise condition). We compared two methods (MOSA-Net vs  Quality-Net and MOSA-Net vs STOI-Net) by performing t-test on the individual average MSE/LCC/SRCC scores of 20 matched pairs (where each individual score was computed from 5 utterances), and found that all the performance differences in Table II are statistically significant, with a p-value less than 0.05.



\begin{table}[t]
\caption{LCC, SRCC, and MSE results of MOSA-Net, Quality-Net, and STOI-Net. *denotes that the performance deference of the two models in each condition is statistically significant.}
\footnotesize
\begin{center}
\resizebox{8.75cm}{!}{
 \begin{tabular}{c||c||c||c||c||c||c} 
 \hline
 \hline
  \multirow{2}{*}{\textbf{Model}} & \multicolumn{3}{c||}{Seen Noises} &  \multicolumn{3}{c} {Unseen Noises}  \\ \cline{2-7}
  
  &\textbf{LCC} & \textbf{SRCC} & \textbf{MSE}&\textbf{LCC} & \textbf{SRCC} & \textbf{MSE}  \\ [0.5ex] \cline{2-7}
 \hline\hline
  \multicolumn{7}{c} {PESQ score prediction} \\
 \hline
Quality-Net \cite{ref_49}&0.964&0.945&0.074&0.957&0.932&\textbf{0.075*}\\ \hline
MOSA-Net&\textbf{0.982*}&\textbf{0.967*}&\textbf{0.040*}&\textbf{0.965*}&\textbf{0.954*}&0.092\\ \hline
 
 \hline
  \multicolumn{7}{c} {STOI score prediction} \\
 \hline
STOI-Net \cite{ref_52}&0.964&0.962&0.002&0.789&0.797&0.016\\ \hline
MOSA-Net&\textbf{0.970*}&\textbf{0.968*}&\textbf{0.001*}&\textbf{0.827*}&\textbf{0.815*}&\textbf{0.015*}\\ \hline

\end{tabular}}
\end{center}
\end{table}
\par

\graphicspath{ {./images/} }
\begin{figure}[t]
\centering
\includegraphics[width=8cm]{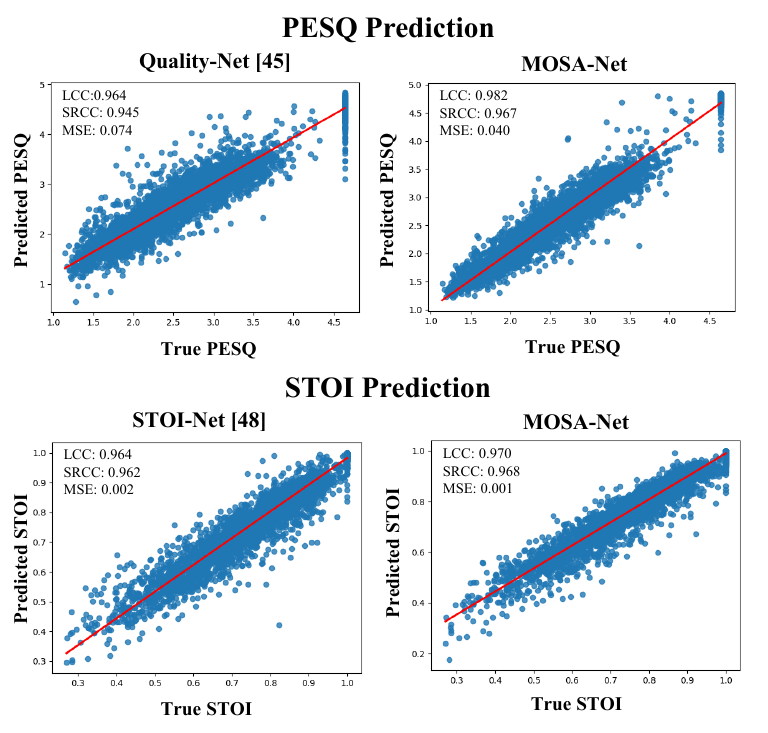} 
\caption{Scatter plots of speech assessment predictions of MOSA-Net, Quality-Net \cite{ref_49}, and STOI-Net \cite{ref_52}.} 
\label{fig:Scatter_single}
\end{figure}

\subsubsection{MOSA-Net with single- and multi-task training}
Next, we aimed to compare the performance of the MOSA-Net with single- and multi-task training criteria. In the previous section, we used a single-task training criterion. Specifically, when the prediction task was the PESQ/STOI, the MOSA-Net was trained using PESQ/STOI labels. In this section, we used multiple assessment targets to train the MOSA-Net, and the model architecture is shown in Fig. 1. The results of single-, double-, and triple-task learning are shown in Table III, where the prediction targets are the PESQ, STOI, and SDI scores, respectively.  

As shown in Table III, in PESQ score prediction, under the seen noise condition, the MOSA-Net models trained with double-task criteria (Q+I and Q+D) yielded better results in terms of all LCC, SRCC, and MSE metrics than MOSA-Net trained with a single-task criterion (Q), and the triple-task criterion (Q+I+D) yielded the best performance. However, for the unseen noise condition, neither double-task nor triple-task criteria led to performance gains. For STOI score prediction, similar trends can be observed in Table III. For SDI score prediction, MOSA-Net (Q+D) and MOSA-Net (I+D) achieved better performance than MOSA-Net (D) in both seen and unseen noise conditions. Additional consideration of PESQ or STOI in model training really helps with SDI score prediction. While additionally considering both PESQ and STOI in model training yielded good results, these results were not the best. We performed t-test on the results in Table III. For each testing condition, the results of the best multi-task model and the corresponding single-task model were compared. For example, for the SRCC metric in PESQ score prediction under the seen noise condition, MOSA-Net (Q+I+D) and MOSA-Net (Q) were compared. From Table III, we can see that when a multi-task model outperformed the corresponding single-task model, the performance improvement (e.g., MOSA-Net (Q+I+D) outperformed MOSA-Net (Q) in PESQ prediction in terms of SRCC: 0.977 vs 0.965) was almost always statistically significant, with a p-value less than 0.05. In contrast, when the best multi-task model performed worse than the corresponding single-task model, there was only one case where the difference in performance was statistically significant, i.e., MOSA-Net (Q+I) vs MOSA-Net (I) in STOI prediction in terms of LCC under the unseen noise condition. Overall, the results in Table III suggest that PESQ, STOI, and SDI predictions are correlated to some extent and that it is beneficial to employ a multi-task learning criterion when training speech assessment models.


\begin{table}[t]
\caption{LCC, SRCC, and MSE results of MOSA-Net trained with single (Q/I/D), double (Q+I/Q+D/I+D), and triple (Q+I+D) metrics for predicting PESQ, STOI, SDI scores under seen and unseen conditions. Q, I, and D denote PESQ, STOI, and SDI scores, respectively. *denotes that the performance deference of the best multi-task model and the single-task model in each condition is statistically significant.}
\footnotesize
\setlength\tabcolsep{4pt}
\begin{center}
 \begin{tabular}{c||c||c||c||c||c||c} 
 \hline
 \hline
  \multirow{2}{*}{\textbf{Model}} & \multicolumn{3}{c||}{Seen Noises} &  \multicolumn{3}{c} {Unseen Noises}  \\ \cline{2-7}
  
  &\textbf{LCC} & \textbf{SRCC} & \textbf{MSE}&\textbf{LCC} & \textbf{SRCC} & \textbf{MSE}  \\ [0.5ex] \cline{2-7}
 \hline\hline
  \multicolumn{7}{c} {PESQ score prediction} \\
 \hline

Q&0.982&0.965&0.043&0.965&\textbf{0.954}&0.092\\ \hline
Q+I&0.987&0.974&0.028&\textbf{0.966}&0.952&\textbf{0.068*}\\ \hline
Q+D&0.986&0.975&0.028&0.956&0.939&0.100\\ \hline
Q+I+D&\textbf{0.988*}&\textbf{0.977*}&\textbf{0.026*}&0.965&0.950&0.075\\ \hline

 \hline
  \multicolumn{7}{c} {STOI score prediction} \\
 \hline
 
I&0.970&0.968&\textbf{0.001}&\textbf{0.827*}&0.815&0.015\\ \hline
Q+I&0.971&0.968&0.002&0.802&0.815&\textbf{0.014}\\ \hline
I+D&0.973&0.968&\textbf{0.001}&0.785&0.792&0.015\\ \hline
Q+I+D&\textbf{0.977*}&\textbf{0.974*}&\textbf{0.001}&0.790&\textbf{0.816}&0.016\\ \hline

 \hline
  \multicolumn{7}{c} {SDI score prediction} \\
 \hline

D&0.883&0.904&0.045&0.826&0.822&0.050\\ \hline
Q+D&0.939&0.947&0.024&\textbf{0.866*}&\textbf{0.866*}&0.038\\ \hline
I+D&\textbf{0.952*}&\textbf{0.955*}&\textbf{0.019*}&0.848&0.826&0.042\\ \hline
Q+I+D&0.947&0.954&0.022&0.850&0.859&\textbf{0.036*}\\ \hline
 
\end{tabular}
\end{center}
\end{table}
\par

\graphicspath{ {./images/} }
\begin{figure}[t]
\centering
\includegraphics[width=8cm]{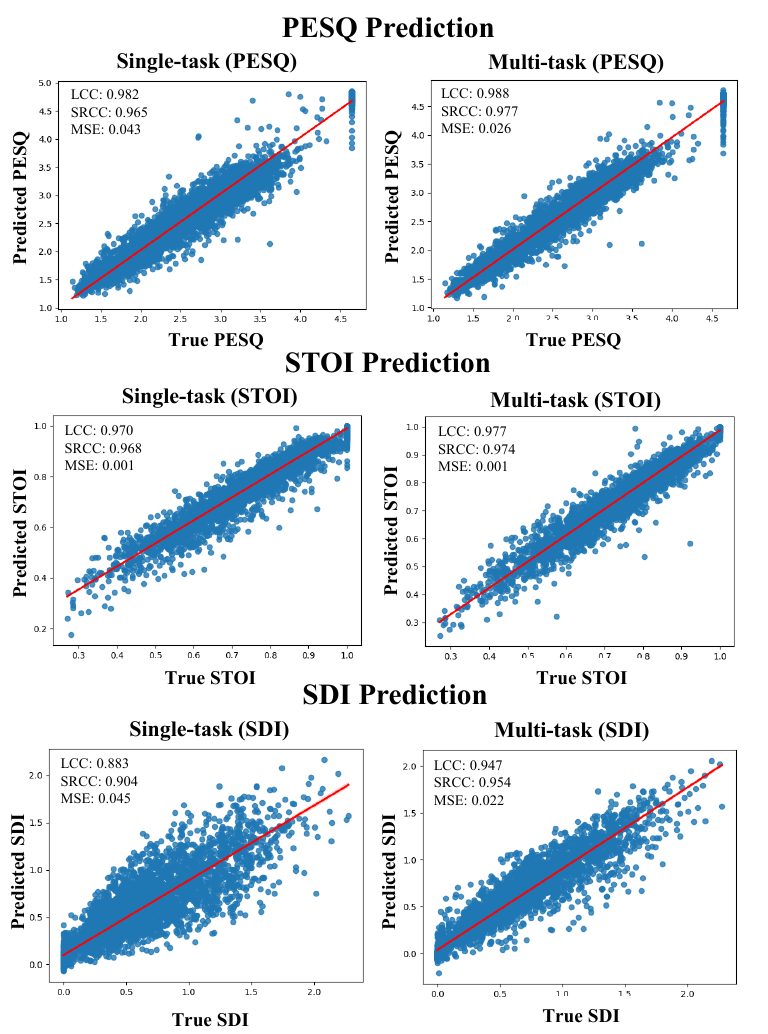} 
\caption{Scatter plots of speech assessment predictions of the single-task and multi-task MOSA-Net models.}
\label{fig:Scatter}
\end{figure}

In addition to quantitative analyses, we conducted qualitative analyses on the MOSA-Net trained with single- and multi-task training criteria. As shown in Fig. 4, the multi-task MOSA-Net could estimate the assessment scores more accurately than the single-task MOSA-Net models, as evidenced by the denser distribution of points along the diagonal. 

To develop a comprehensive analysis, we visualized the hidden layer representation of the MOSA-Net with single-task and triple-task learning. We extracted the output of the attention layer from each of the models. In addition, we present the scatter plots of MOSA-Net trained with single- and multi-task criteria in Fig. 5. From the figure, the representations of the single-task MOSA-Net trained with individual PESQ, STOI, and SDI values yielded different patterns when predicting the individual metrics (the PESQ, STOI, and SDI). This shows that the MOSA-Net trained with a distinct metric is learned to focus on particular regions. By contrast, as shown in Fig. 6, the multi-task MOSA-Net that was trained simultaneously on three assessment metrics yielded different visualization results. Unlike the single-task models, the multi-task MOSA-Net model yielded a similar pattern in each of the branches. Therefore, it may further suggest that the MOSA-Net aims to share useful representations and achieve more general weights by optimally considering all metrics. 

\subsubsection{Comparison with another multi-task method}
\graphicspath{ {./images/} }
\begin{figure}[t]
\centering
\includegraphics[width=7cm]{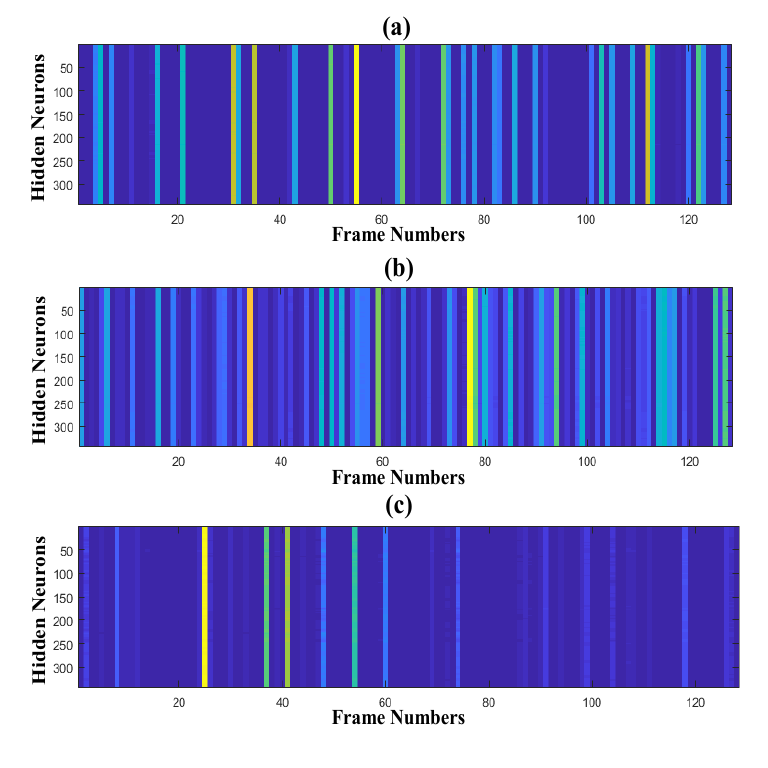} 
\caption{Latent representations of a speech utterance at the attention layer of the single-task MOSA-Net (a) PESQ, (b) STOI, and (c) SDI. The horizontal and vertical axes denote the frame index and attention weight, respectively.}
\label{fig:LR1}
\end{figure}

\graphicspath{ {./images/} }
\begin{figure}[ht]
\centering
\includegraphics[width=7cm]{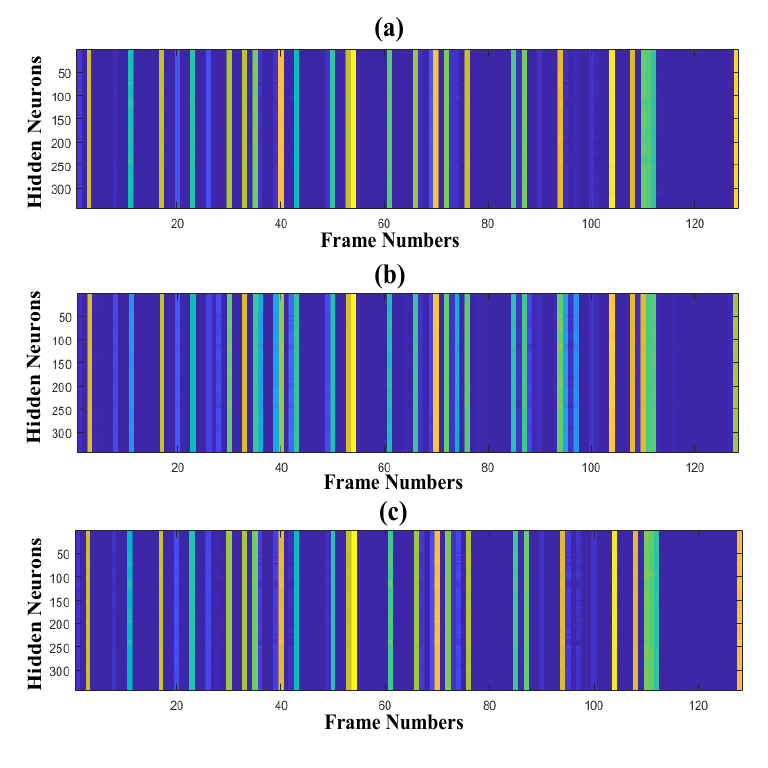} 
\caption{Latent representations of a speech utterance at the attention layer of the multi-task MOSA-Net (a) PESQ, (b) STOI, and (c) SDI. The horizontal and vertical axes denote the frame index and attention weight, respectively.}
\label{fig:LR2}
\end{figure}

\begin{table}[t]
\caption{LCC, SRCC, and MSE results of MOSA-Net and AMSA.}
\footnotesize
\setlength\tabcolsep{4pt}
\begin{center}
 \begin{tabular}{c||c||c||c||c||c||c} 
 \hline
 \hline
  \multirow{2}{*}{\textbf{Model}} & \multicolumn{3}{c||}{Seen Noises} &  \multicolumn{3}{c} {Unseen Noises}  \\ \cline{2-7}
  
  &\textbf{LCC} & \textbf{SRCC} & \textbf{MSE}&\textbf{LCC} & \textbf{SRCC} & \textbf{MSE}  \\ [0.5ex] \cline{2-7}
 \hline\hline
  \multicolumn{7}{c} {PESQ score prediction} \\
 \hline

AMSA \cite{ref_56}&0.985&0.973&0.031&0.962&0.946&0.080\\ \hline
MOSA-Net &\textbf{0.988}&\textbf{0.977}&\textbf{0.026}&\textbf{0.965}&\textbf{0.950}&\textbf{0.075}\\ \hline

 \hline
  \multicolumn{7}{c} {STOI score prediction} \\
 \hline
 
AMSA \cite{ref_56}&0.975&0.973&\textbf{0.001}&0.783&0.794&0.018\\ \hline
MOSA-Net &\textbf{0.977}&\textbf{0.974}&\textbf{0.001}&\textbf{0.790}&\textbf{0.816}&\textbf{0.016}\\ \hline

 \hline
  \multicolumn{7}{c} {SDI score prediction} \\
 \hline

AMSA \cite{ref_56}&0.929&0.942&0.029&0.835&0.847&0.039\\ \hline
MOSA-Net &\textbf{0.947}&\textbf{0.954}&\textbf{0.022}&\textbf{0.850}&\textbf{0.859}&\textbf{0.036}\\ \hline
 
\end{tabular}
\end{center}
\end{table}
\par

In this section, we compare the performance of the MOSA-Net with that of another multi-task speech assessment model, namely, attention enhanced multi-task speech assessment (AMSA) \cite{ref_56}. Specifically, we compared two different strategies for constructing the objective function. In our proposed work, we combined the estimated loss from the utterance- and frame-level scores to define the objective function. By contrast, AMSA uses the regression loss based on the utterance-level score and the classification loss based on the classification-aided model to define the objective function. For a fair comparison, the same model architecture with the same number of assessment targets was used in both systems. When training the AMSA system, we followed the same parameter as defined in \cite{ref_56} to adjust the classification-aided model.

As shown in Table IV, MOSA-Net consistently outperformed AMSA in PESQ, STOI, and SDI predictions across all LCC, SRCC, and MSE metrics. Therefore, these evaluation results demonstrate the benefit of combining the utterance-level score and the frame-level score to form the objective function. 

\subsubsection{MOSA-Net with cross-domain features}
In this section, we investigate the effectiveness of different acoustic features on the performance of MOSA-Net \footnote{https://github.com/dhimasryan/MOSA-Net-Cross-Domain} and whether the combination of multiple acoustic features can lead to more accurate predictions for MOSA-Net. In addition to PS features, which have been used in the previous experiments, MOSA-Net employed complex features (termed complex), learnable filter banks (termed LFB features), and the output of a self-supervised learned model (termed SSL features). The goals of using these three types of features are as follows: (1) Complex features can reserve the phase information; (2) LFB features can retain the raw-waveform information more completely; (3) SSL features can exploit the context-information of phones. For (1), we used real and imaginary (RI) spectrograms. For (2), we used SincNet \cite{sincnet} as the learnable feature extraction model. For (3), we used two self-supervised learned models, namely wav2vec 2.0 \cite{wav2vec} and HuBERT \cite{hubert}, to generate the SSL features. The corresponding features are termed SSL(W2V) and SSL(Hub), respectively. The results of MOSA-Net using PS, Complex, LFB, SSL(W2V), and SSL(Hub) features are shown in Table V.

\begin{table}[t]
\caption{LCC, SRCC, and MSE results of MOSA-Net using different input features.}
\footnotesize
\begin{center}
 \begin{tabular}{c||c||c||c||c||c||c} 
 \hline
 \hline
  \multirow{2}{*}{\textbf{Model}} & \multicolumn{3}{c||}{Seen Noises} &  \multicolumn{3}{c} {Unseen Noises}  \\ \cline{2-7}
  
  &\textbf{LCC} & \textbf{SRCC} & \textbf{MSE}&\textbf{LCC} & \textbf{SRCC} & \textbf{MSE}  \\ [0.5ex] \cline{2-7}
 \hline\hline
  \multicolumn{7}{c} {PESQ score prediction} \\
 \hline

PS&\textbf{0.988}&\textbf{0.977}&\textbf{0.026}&\textbf{0.965}&0.950&\textbf{0.075}\\ \hline
Complex&0.985&0.975&0.031&\textbf{0.965}&\textbf{0.951}&0.081\\ \hline
LFB&0.981&0.971&0.040&0.957&0.942&0.091\\ \hline
SSL(W2V)&0.984&0.972&0.033&0.961&0.947&0.084\\ \hline
SSL(Hub)&0.981&0.967&0.041&0.954&0.933&0.088\\ \hline
 
 \hline
  \multicolumn{7}{c} {STOI score prediction} \\
 \hline
 
PS&0.977&0.974&\textbf{0.001}&0.790&0.816&0.016\\ \hline
Complex&0.976&0.974&\textbf{0.001}&0.765&0.794&\textbf{0.014}\\ \hline
LFB&0.972&0.970&\textbf{0.001}&0.778&0.787&0.016\\ \hline
SSL(W2V)&0.970&0.968&0.002&0.804&0.820&0.017\\ \hline
SSL(Hub)&\textbf{0.980}&\textbf{0.978}&\textbf{0.001}&\textbf{0.807}&\textbf{0.821}&0.015\\ \hline

 \hline
  \multicolumn{7}{c} {SDI score prediction} \\
 \hline

PS&\textbf{0.947}&\textbf{0.954}&\textbf{0.022}&\textbf{0.850}&\textbf{0.859}&\textbf{0.036}\\ \hline
Complex&0.945&0.953&0.023&0.818&0.839&0.047\\ \hline
LFB&0.936&0.944&0.025&0.827&0.834&0.056\\ \hline
SSL(W2V)&0.890&0.918&0.043&0.822&0.836&0.074\\ \hline
SSL(Hub)&0.935&0.952&0.026&0.842&0.830&0.068\\ \hline
 
\end{tabular}
\end{center}
\end{table}

\begin{table}[t]
\caption{LCC, SRCC, and MSE results of MOSA-Net using cross-domain features.}
\footnotesize
\begin{center}
\setlength\tabcolsep{3pt}
 \begin{tabular}{c||c||c||c||c||c||c} 
 \hline
 \hline
  \multirow{2}{*}{\textbf{Model}} & \multicolumn{3}{c||}{Seen Noises} &  \multicolumn{3}{c} {Unseen Noises}  \\ \cline{2-7}
  
  &\textbf{LCC} & \textbf{SRCC} & \textbf{MSE}&\textbf{LCC} & \textbf{SRCC} & \textbf{MSE}  \\ [0.5ex] \cline{2-7}
 \hline\hline
  \multicolumn{7}{c} {PESQ score prediction} \\
 \hline

PS+SSL(Hub)&\textbf{0.991}&\textbf{0.981}&\textbf{0.020}&0.968&\textbf{0.957}&\textbf{0.066}\\ \hline
Complex+SSL(Hub)&0.990&0.979&0.023&0.968&0.956&0.084\\ \hline
LFB+SSL(Hub) &0.989&0.978&0.024&0.963&0.951&0.085\\ \hline
PS+LFB+SSL(Hub) &0.990&0.980&0.021&\textbf{0.969}&\textbf{0.957}&0.070\\ \hline
Complex+LFB+SSL(Hub)&0.990&0.980&0.022&0.967&0.956&0.081\\ \hline

 \hline
  \multicolumn{7}{c} {STOI score prediction} \\
 \hline
 
PS+SSL(Hub)&\textbf{0.989}&0.985&\textbf{0.001}&0.814&0.820&\textbf{0.016}\\ \hline
Complex+SSL(Hub)&\textbf{0.989}&\textbf{0.986}&\textbf{0.001}&0.826&0.828&0.015\\ \hline
LFB+SSL(Hub) &0.986&0.984&\textbf{0.001}&0.834&0.834&0.022\\ \hline
PS+LFB+SSL(Hub) &0.985&0.984&\textbf{0.001} &\textbf{0.836}&\textbf{0.839}&0.017\\ \hline
Complex+LFB+SSL(Hub)&\textbf{0.989}&0.985&\textbf{0.001}&0.831&0.826&0.016\\ \hline

 \hline
  \multicolumn{7}{c} {SDI score prediction} \\
 \hline

PS+SSL(Hub)&0.961&0.966&0.016&0.878&0.866&0.044\\ \hline
Complex+LFB&0.942&0.950&0.023&0.839&0.857&0.048\\ \hline
Complex+SSL(Hub)&\textbf{0.971}&\textbf{0.973}&\textbf{0.012}&0.890&0.888&0.037\\ \hline
LFB+SSL(Hub) &0.964&0.970&0.014&0.851&0.836&0.060\\ \hline
PS+LFB+SSL(Hub) &0.964&0.967&0.015&0.878&0.872&0.045\\ \hline
Complex+LFB+SSL(Hub)&0.969&0.971&0.012&\textbf{0.895}&\textbf{0.899}&\textbf{0.033}\\\hline
\end{tabular}
\end{center}
\end{table}
\par

\graphicspath{ {./images/} }
\begin{figure}[t]
\centering
\includegraphics[width=8cm]{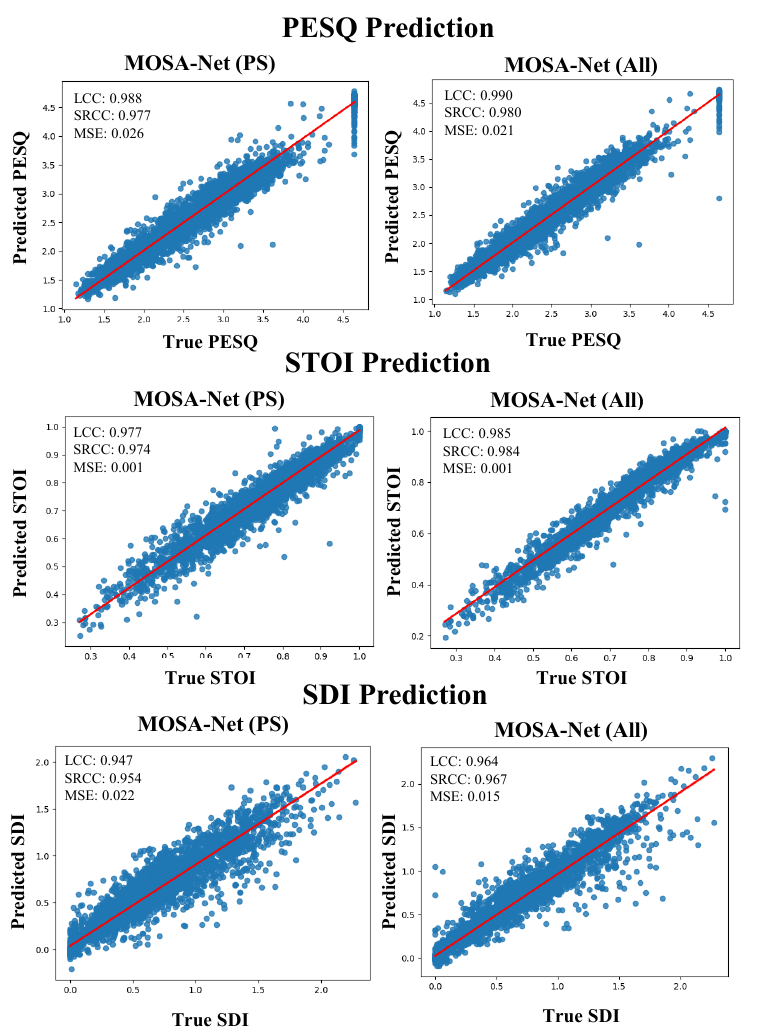} 
\caption{Scatter plots of speech assessment predictions of the MOSA-Net with the single-domain (PS) and cross-domain (PS+LFB+SSL(Hub)) features.} 
\label{fig:Scatter_multi}
\end{figure}

\par
As shown in Table V, the PS features tended to achieve slightly better performance than the other features when estimating the PESQ. By contrast, in assessing the STOI score, the SSL(Hub) features achieved better performance in both the seen and unseen environments. Meanwhile, in assessing the SDI score, the PS features achieved better performances in the seen and unseen environments, respectively. Hence, it is indicated that these acoustic features have different and complementary information for speech assessment. In addition, by considering phase information, the Complex features can reserve useful information that is particularly more useful when conducting assessment evaluations in seen environments. Because SSL(Hub) generally provides better performance than SSL(W2V), SSL(Hub) is used as the representative SSL features in the following discussion.
\par

Next, we further investigated the MOSA-Net that combines cross-domain features as input. As shown in Fig. 1, the STFT and learnable neural network (SincNet in this study) were applied to the speech waveform to obtain the PS/Complex and LFB features, which were then used as the input to the MOSA-Net. For the SSL(W2V) and SSL(Hub) features, the speech waveform was processed by the wav2vec 2.0 and HuBERT models, respectively, and the latent representations were input to the middle layer of the MOSA-Net model. The results of the MOSA-Net with different combinations of acoustic features are shown in Table VI. 

\par
Comparing the results in Table VI and Table V, the benefits of incorporating cross-domain features to train the MOSA-Net model are evident. For example, the combination of Complex and SSL(Hub), denoted as Complex+SSL(Hub) in Table VI, consistently outperformed the individual Complex and SSL(Hub) in PESQ, STOI, and SDI predictions in both the seen and unseen environments. Furthermore, Table VI shows that the Complex+Hub features achieved the best performance among all combinations for STOI and SDI predictions in the seen environments. Finally, the combination of three acoustic features, namely Complex+LFB+SSL(Hub)/PS+LFB+SSL(Hub), consistently achieved better performance in the unseen environments as compared with Complex+SSL(Hub) / PS+SSL(Hub).  We also present the scatter plots of predictions of the MOSA-Net with the single-domain (PS) and cross-domain (PS+LFB+SSL(Hub)) features in Fig. 7. From the figure, the MOSA-Net with the cross-domain features can achieve a more accurate estimation than the MOSA-Net with the single-domain features. In addition, the t-test shows that the improvements of the MOSA-Net with cross-domain (PS+LFB+SSL(Hub)) features over the MOSA-Net with single-domain (PS) features in PESQ, STOI, and SDI score predictions for both seen and unseen environments in terms of LCC, SRCC, and MSE are all statistically significant, with a p-value less than 0.05. The results confirm the benefit of cross-domain features, which provide more complete information for the speech assessment model.

  





\begin{table}[t]
\caption{LCC, SRCC, and MSE results of MOSA-Net tested on the TIMIT dataset.}
\footnotesize
\begin{center}
 \begin{tabular}{c||c||c||c} 
 \hline
 \hline
 \textbf{Model} &\textbf{LCC} & \textbf{SRCC} & \textbf{MSE}  \\ [0.5ex] \cline{2-4}
 \hline\hline
  \multicolumn{4}{c} {PESQ Score Prediction} \\
 \hline
AMSA \cite{ref_56}&0.728&0.673&0.765\\ \hline
MOSA-Net&0.754&0.710&0.654\\ \hline
MOSA-Net(Cross-Domain)&\textbf{0.960}&\textbf{0.948}&\textbf{0.111}\\ \hline
 \hline
  \multicolumn{4}{c} {STOI Score Prediction} \\
 \hline
AMSA \cite{ref_56}&0.701&0.584&0.015\\ \hline
MOSA-Net&0.746&0.608&0.011\\ \hline
MOSA-Net(Cross-Domain)&\textbf{0.920}&\textbf{0.936}&\textbf{0.004}\\ \hline
 \hline
   \multicolumn{4}{c} {SDI Score Prediction} \\
 \hline
AMSA \cite{ref_56}&0.873&0.678&\textbf{0.042}\\ \hline
MOSA-Net&0.859&0.643&0.047\\ \hline
MOSA-Net(Cross-Domain)&\textbf{0.895}&\textbf{0.901}&0.051\\ \hline
 \hline
 \hline

\end{tabular}
\end{center}
\end{table}

\graphicspath{ {./images/} }
\begin{figure}[ht]
\centering
\includegraphics[width=9cm]{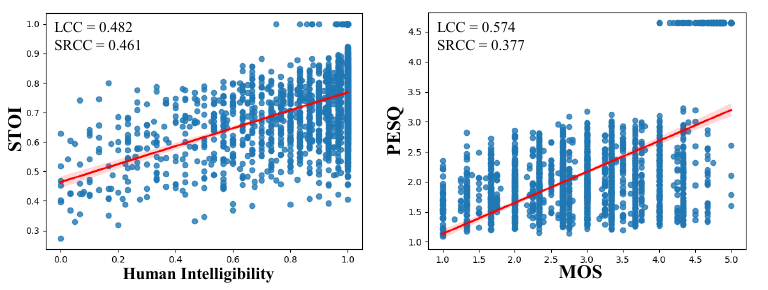} 
\caption{Scatter plots between subjective intelligibility and STOI (left), and subjective quality and PESQ (right).}
\label{fig:Scatter_multi}
\end{figure}


\subsubsection{MOSA-Net tested on the TIMIT dataset}
In this section, we aim to analyze the generalization ability of MOSA-Net by testing the models trained on the WSJ dataset on another standard English dataset, TIMIT \cite{TIMIT}. The testing noisy set was obtained by injecting the same set of noises used in the WSJ task at eight SNR levels (from -10 dB to 25 dB with a step of 5 dB) to the original clean utterances, and the enhanced set was obtained by applying the same BLSTM-based SE model used in the WSJ task to the noisy utterances. Therefore, the evaluation experiment was performed under the seen noise condition. In total, we randomly selected 750 noisy utterances, 750 enhanced utterances, and 500 clean utterances to test the assessment models, including AMSA [46], MOSA-Net using PS features (denoted as MOSA-Net), and MOSA-Net using cross-domain features (denoted as MOSA-Net Cross-Domain). As shown in Table VII, MOSA-Net consistently outperforms AMSA in PESQ and STOI predictions while slightly underperforms AMSA in SDI prediction, but MOSA-Net Cross-Domain achieves the best performance in almost all testing conditions except for the MSE metric for SDI prediction. The results show that by using cross-domain features, MOSA-Net(Cross-Domain) is obviously more robust across datasets, with clear advantages over AMSA and MOSA-Net using single features. The t-test confirmed that the improvements of MOSA-Net(Cross-Domain) over AMSA in 8 out of 9 testing conditions are all statistically significant, with a p-value less than 0.05. Overall, the results confirm the effectiveness of using cross-domain features to improve generalization to new datasets.

\begin{table}[t]
\caption{LCC, SRCC, and MSE results of MOSA-Net for Human Listening Test Prediction.}
\footnotesize
\begin{center}
 \begin{tabular}{c||c||c||c} 
 \hline
 \hline
 \textbf{Model} &\textbf{LCC} & \textbf{SRCC} & \textbf{MSE}  \\ [0.5ex] \cline{2-4}
 \hline\hline
  \multicolumn{4}{c} {MOS Prediction}
\\ \hline
CSIG\cite{hu2007evaluation}&0.555&0.453&-\\\hline
CBAK\cite{hu2007evaluation}&0.545&0.343&-\\\hline
COVL\cite{hu2007evaluation}&0.556&0.450&-\\\hline
MOSNet \cite{ref_52}&0.724&0.656&0.489\\ \hline
MOS-SSL \cite{ssl-mos}&0.787&0.746&0.440\\ 
 \hline
MOSA-Net(WSJ)&0.535&0.371&1.636\\ \hline
MOSA-Net(Scratch)&0.777&0.724&0.411\\ \hline
MOSA-Net(Adapt)&0.795&0.742&0.389\\ \hline
MOSA-Net(Scratch)$_{\rm FT-SSL}$&0.804&0.758&0.360\\ \hline
MOSA-Net(Adapt)$_{\rm FT-SSL}$&\textbf{0.805}&\textbf{0.763}&\textbf{0.356}\\ \hline
 \hline
  \multicolumn{4}{c} {Speech Intelligibility Prediction} \\
 \hline
ESTOI \cite{estoi} &0.461&0.465&0.162\\ \hline
MOSNet \cite{ref_52}&0.658&0.607&0.027\\ \hline
MOS-SSL \cite{ssl-mos}&0.760&0.655&0.024\\ \hline
MOSA-Net(WSJ)&0.385&0.378&0.056\\ \hline
MOSA-Net(Scratch)&0.740&0.698&0.023\\ \hline
MOSA-Net(Adapt)&0.756&0.702&0.021\\ \hline
MOSA-Net(Scratch)$_{\rm FT-SSL}$&0.796&0.712&0.018\\ \hline
MOSA-Net(Adapt)$_{\rm FT-SSL}$&\textbf{0.807}&\textbf{0.730}&\textbf{0.017}\\ \hline
 \hline

\end{tabular}
\end{center}
\end{table}
\subsubsection{Adapting MOSA-Net to predict human subjective ratings}
Collecting subjective scores is time consuming and expensive because multiple subjects are required for listening tests. However, compared with models that predict objective evaluation scores, it is much more challenging to train a model to predict subjective evaluation scores that are highly correlated with human subjective ratings because human subjective ratings vary greatly, as shown in some previous work \cite{ref_51, wav2vec_mos,mbnet_mos}. Therefore, there is an urgent need for a method that can effectively and efficiently train a model that replaces subjective evaluation. In the experiments, we used the utterances and corresponding subjective quality and intelligibility scores in the TMHINT-QI dataset \cite{TMINT-QI}\footnote{TMHINT-QI dataset: http://gofile.me/6PGhz/4U6GWaOtY; TMHINT-QI dataset description: https://github.com/yuwchen/InQSS}. The dataset includes clean, noisy, and enhanced utterances of five SE methods (namely Karhunen-Loeve transform (KLT) \cite{klt}, minimum-mean square error (MMSE) \cite{mmse}, fully convolutional network (FCN) \cite{FCN}, deep denoising autoencoder (DDAE) \cite{DDAE}, and transformer-based SE \cite{Trans}). 226 subjects participated in the listening test (subjective quality and intelligibility)\footnote{Written informed consent approved by the Academia Sinica Institutional Review Board for this study was obtained from each participant before conducting the experiment.}. Most utterances were evaluated by one subject, but some utterances were evaluated by more than one subject. In \cite{hu2007evaluation}, it is presented that there are two setups for speech metric predictions: the corresponding prediction targets can be either average testing-condition-specific rating scores from multiple subjects or rating scores from an individual or single subject. In this study, we followed the later setup. The quality score ranges from 1 to 5, where a higher score indicates higher perceived quality. The intelligibility score ranges from 0 to 1, indicating the percentage of recognizable characters. We used 1,900 utterances evaluated by multiple subjects for testing, and selected 15,000 utterances evaluated by one subject for training. For each test utterance, the average score was used as the ground-truth score. It is noteworthy that the training and test utterances do not overlap. Besides, the number of clean, noisy, and enhanced utterances of five SE models is roughly balanced.

First, we investigate whether MOSA-Net can be adapted to a new model for estimating human subjective ratings with a limited amount of training data. We used scatter plots to show the correlations between human-listening test scores (intelligibility and MOS) and objective test scores (STOI and PESQ). We drew two scatter plots in Fig.8 (left panel: subjective intelligibility vs STOI; right panel: subjective MOS vs PESQ). We used all the test utterances to conduct the evaluation. The average MOS scores for individual speech utterances appear to be distributed only over a limited number of values because each subject rated an utterance as 1, 2, 3, 4, or 5. Considering each utterance consists of 10 Chinese characters, the human-assessed recognition rate is either 0.1, 0.2, 0.3, 0.4, 0.5, 0.6, 0.7, 0.8, 0.9 or 1.0. Therefore, the average speech intelligibility scores for individual speech utterances also appear to be distributed only over a limited number of values. We also observe that the PESQ (or STOI) scores of utterances with the same average MOS (or intelligibility) score are diverse. From the two scatter plots, we note that human-listening test results (intelligibility and MOS) have moderate correlations with the corresponding objective scores (STOI and PESQ). The correlation between speech intelligibility and STOI is evaluated as 0.482 (LCC),and 0.461 (SRCC), while the correlation between MOS and PESQ is evaluated as 0.574 (LCC),and 0.377 (SRCC). These findings motivated us to use a model adaptation strategy to adapt the MOSA-Net model pre-trained on objective scores to a new model that can predict human subjective ratings.

In the next experiment, we used two state-of-the-art speech assessment models as our baselines: (1) MOSNet: a model that is based on a CNN-BLSTM architecture for predicting MOS scores \cite{ref_51}; (2) MOS-SSL: a model that uses features from fine-tuned wav2vec 2.0 to predict MOS scores \cite{ssl-mos}. Both models were trained on the TMHINT-QI dataset with a single-task criterion to predict the quality or intelligibility score separately. MOSNet was trained from scratch. For MOS-SSL, its linear output layer was trained from scratch while the wav2vec module was fine-tuned. In addition, several intrusive speech quality prediction approaches such as CSIG: MOS predictor of signal distortion, CBAK: MOS predictor of background-noise, and COVL: MOS
predictor of overall signal quality \cite{hu2007evaluation} were selected for speech quality prediction. Along with that the intrusive speech intelligibility prediction approach called ESTOI \cite{estoi} was used for speech intelligibility prediction.
We compared five MOSA-Net models: (1) MOSA-Net(WSJ): the best model trained on WSJ (i.e., PS+LFB+SSL(Hub) in Table VI); (2) MOSA-Net(Scratch): a model that is trained from scratch on the TMHINT-QI dataset with the same configuration as the best MOSA-Net; (3) MOSA-Net(Adapt): a model that is adapted from MOSA-Net(WSJ) using the TMHINT-QI dataset; (4) MOSA-Net(Scratch)$_{\rm FT-SSL}$: same as MOSA-Net(Scratch), except that the SSL model was the one that was fine-tuned in MOS-SSL; (5) MOSA-Net(Adapt)$_{\rm FT-SSL}$: same as MOSA-Net(Adapt), except that the SSL model was the one that was fine-tuned in MOS-SSL. The learning rate for adaptation was set to 0.00005 (half the learning rate used in MOSA-Net training).

The results are listed in Table VIII. Obviously, due to data mismatch and the gap between the PESQ/STOI metrics and the subjective quality/intelligibility scores, MOSA-Net(WSJ) could not yield satisfactory performance (especially intelligibility score prediction). In contrast, MOSA-Net(Scratch) and MOSA-Net(Adapt) performed notably better than MOSA-Net(WSJ), and several well-known intrusive evaluation metrics (CSIG, CBAK, COVL, and ESTOI). Meanwhile, MOSA-Net(Adapt) was superior to MOSA-Net(Scratch) in both speech quality and intelligibility predictions. It is also noted that with the fine-tuned SSL features, MOSA-Net(Scratch)$_{\rm FT-SSL}$ outperformed MOSA-Net(Scratch) and MOSA-Net(Adapt), and that by utilizing the prior weight information, MOSA-Net(Adapt)$_{\rm FT-SSL}$ achieved the best performance among all MOSA-Net models. Furthermore, MOSA-Net(Adapt)$_{\rm FT-SSL}$ outperformed the two well-known baseline models, MOSNet and MOS-SSL. The t-test shows that the improvements of MOSA-Net(Adapt)$_{\rm FT-SSL}$ over MOS-SSL in both MOS and subjective intelligibility predictions in all metrics are statistically significant, with a p-value less than 0.05. The results confirm that MOSA-Net(WSJ) can serve as a pre-trained model to be adapted to a new model to predict human subjective ratings. Furthermore, fine-tuning the SSL model can provide more meaningful feature representations, leading to improved prediction performance.

\subsection{Experiments of SE with assessment information}
In this section, we evaluate the QIA-SE system that incorporates the knowledge from the MOSA-Net model for improving the SE performance. To date, several methods have been proposed to incorporate the knowledge from the speech assessment models into an SE system, e.g., \cite{zezario2019specialized, ref_54}. We intend to compare the proposed QIA-SE system with the comparative SE systems. We tested the proposed QIA-SE system on two SE datasets, namely the WSJ corpus and the Taiwan Mandarin version of the Hearing in Noise Test (TMHINT) dataset \cite{ref_67}. PESQ, STOI, signal distortion (CSIG) \cite{hu2007evaluation}, and segmental signal to noise ratio improvement (SSNRI) scores were used to evaluate the SE performance.

\subsubsection{Experiments on the WSJ dataset}

\par
We used the same 37,416 noisy-clean pairs in Section IV.A.1 to form the training set. From the test set of WSJ, we used four seen noise types (i.e., white, engine, bell, and traffic) and four unseen noise types (i.e., car, pink, street, and babble) to prepare 330 seen noisy test utterances and 330 unseen noisy test utterance at six SNR levels (i.e., -10, -5, 0, 5, 10, and 15 dB). All training and test utterances were converted to 257-dimensional log-power-spectra (LPS) features with a Hamming window of 32 ms, a hop of 16 ms, and a 512-point STFT.
\par
The baseline SE system was built with a CNN model \cite{ref_54}, which comprised 12 convolutional layers, followed by a fully connected layer consisting of 128 neurons. Each convolutional layer contained four channels \{16, 32, 64, 128\} with three types of strides \{1, 1, 3\} in each channel. Two comparative systems, namely specialized speech enhancement model selection (SSEMS) \cite{zezario2019specialized} and zero-shot model selection ZMOS \cite{ref_54}, were constructed to evaluate the effectiveness of the proposed QIA-SE system. In SSEMS, multiple component SE models were prepared, with each model characterizing a particular noisy-clean mapping. Subsequently, a speech assessment model was incorporated to select the most suitable component model based on the estimated PESQ score. In ZMOS, the latent representation of a speech assessment model was used to prepare multiple component models in the offline stage. In the online process, the noisy speech was input into the speech assessment model to obtain the latent representation, which was then used to select the most suitable component model to perform SE. By contrast, the proposed QIA-SE system directly incorporates the latent representation into the hidden layer of the SE model, and is thus a speech-assessment-aware SE system. The enhancement results in terms of PESQ, STOI, CSIG, and SSNRI for the SSEMS, ZMOS, and QIA-SE models are shown in Table IX. Note that the baseline CNN-based SE model and the SSEMS, ZMOS, and QIA-SE models were all implemented on the same CNN architecture.

\begin{table}[t]
\caption{The average SE performance (PESQ, STOI, CSIG, SSNRI) in seen and unseen environments.}
\footnotesize
\begin{center}
\setlength\tabcolsep{4pt}
 \begin{tabular}{c||c||c||c||c||c||c} 
 \hline
 \hline
   \multicolumn{2}{c||} {See}  &  \textbf{Noisy} & \textbf{CNN} & \textbf{SSEMS}  & \textbf{ZMOS} & \textbf{QIA-SE}  \\ [0.5ex] 
\hline   
 \multicolumn{7}{c} {PESQ evaluation} \\
 \cline{1-7}
\multicolumn{2}{c||}{\textbf{Seen}}&2.211&2.652&2.675&2.678&\textbf{2.953}\\
\hline       
\multicolumn{2}{c||}{\textbf{Unseen}}&2.118&2.478&2.484&2.507&\textbf{2.658}\\
 \cline{1-7}
 \multicolumn{7}{c} {STOI evaluation} \\
 \cline{1-7}
\multicolumn{2}{c||}{\textbf{Seen}}&0.830&0.850&0.851&0.851&\textbf{0.868}\\
\hline       
\multicolumn{2}{c||}{\textbf{Unseen}}&0.799&0.820&0.820&0.821&\textbf{0.828}\\
 \cline{1-7}
 \multicolumn{7}{c} {CSIG evaluation} \\
 \cline{1-7}
\multicolumn{2}{c||}{\textbf{Seen}}&1.790&2.091&2.089&2.060&\textbf{2.169}\\
\hline       
\multicolumn{2}{c||}{\textbf{Unseen}}&1.989&2.238&2.227&2.207&\textbf{2.312}\\
 \cline{1-7}
 \multicolumn{7}{c} {SSNRI evaluation} \\
 \cline{1-7}
\multicolumn{2}{c||}{\textbf{Seen}}&-&4.135&4.541&3.640&\textbf{5.516}\\
\hline       
\multicolumn{2}{c||}{\textbf{Unseen}}&-&3.178&3.506&2.683&\textbf{5.731}\\
\hline
\end{tabular}
\end{center}
\end{table}

\par
From Table IX, we first note that both SSEMS and ZMOS outperformed the baseline CNN model. Next, the proposed QIA-SE model significantly outperformed SSEMS and ZMOS in both the seen and unseen environments. In terms of STOI, similar trends were observed, i.e., QIA-SE outperformed the other SE models. In terms of CSIG and SSNRI, QIA-SE also achieved the best performance among all SE systems in both seen and unseen environments. The results confirmed the benefits of applying the speech assessment model as a supportive tool for the main SE task. It is noteworthy that SSEMS and ZMOS adopted deep learning-based speech assessment to prepare multiple component models offline and selected the best one online, where additional models and selection computations are required. By contrast, QIA-SE directly incorporates the latent representation of the assessment model. The results suggest that directly combining the assessment model into the SE system may be a more feasible and hardware-friendly approach.

\subsubsection{Experiments on the TMHINT dataset}
In this experiment, we used the TMHINT dataset to evaluate the proposed QIA-SE model to fulfill three objectives: (1) to further verify the effectiveness of the QIA-SE model on a different SE task (from an English corpus to a Mandarin corpus); (2) to confirm the effectiveness of speech assessment codes across corpora in different languages; and (3) to verify the correlation of the SE performance with the MOSA-Net trained with different training criteria (single-task and multi-task learning). The training set comprised of 1,200 utterances recorded by three male and three female speakers (each speaker provided 200 utterances). We used 100 types of noises \cite{ref_64} to generate 36,000 noisy training utterances at 31 SNR levels (from -10 dB to 20 dB, with a step of 1 dB). The test set comprised 120 utterances recorded by another two speakers (one male and one female). We used four seen noise types (i.e., white, engine, bell, and traffic) and four unseen noise types (i.e., car, pink, street, and babble) to generate 120 seen noisy test utterances and 120 unseen noisy test utterances at six SNR levels (i.e., -10, -5, 0, 5, 10, and 15 dB).

\graphicspath{ {./images/} }
\begin{figure}[t]
\centering
\includegraphics[width=7.5cm]{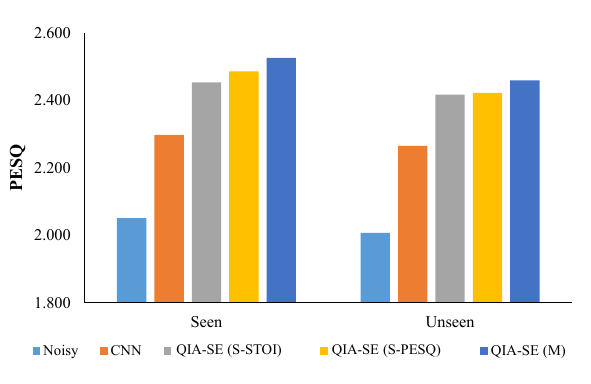} 
\caption{SE performance (PESQ) of Noisy, CNN, QIA-SE (S-PESQ), QIA-SE(S-STOI), and QIA-SE(M).}
\label{fig:PESQ}
\end{figure}

\graphicspath{ {./images/} }
\begin{figure}[t]
\centering
\includegraphics[width=7.5cm]{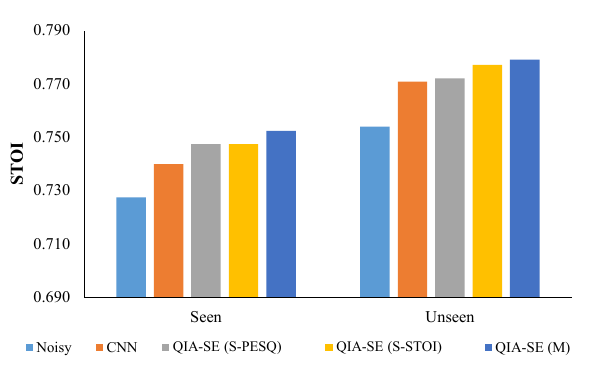} 
\caption{SE performance (STOI) of Noisy, CNN, QIA-SE (S-PESQ), QIA-SE(S-STOI), and QIA-SE(M).} 
\label{fig:STOI}
\end{figure}

\par
Similar to the previous experiments, we used the same CNN architecture to develop all SE systems. We denote the QIA-SE with the MOSA-Net trained with single-task and multi-task criteria as QIA-SE(S) and QIA-SE(M), respectively. The MOSA-Net was constructed based on the best model configuration. The PESQ and STOI results under the seen and unseen noise conditions are shown in Figs. 9 and 10, respectively. QIA-SE(S-PESQ) and QIA-SE(S-STOI) indicate that the PESQ and STOI scores were used to train the single-task MOSA-Net, respectively.
\par
As shown in Figs. 9 and 10, we note that QIA-SE(S-PESQ), QIA-SE(S-STOI), and QIA-SE(M) outperformed the baseline CNN model, whereas QIA-SE(M) achieved better performance than the other two QIA-SE models. The results again confirmed the effectiveness of QIA-SE, which leverages the speech assessment model to attain better SE capability. Meanwhile, as shown in Table III, the multi-task learning criterion allows the MOSA-Net to more accurately predict the speech assessment scores. The results in Figs. 9 and 10 show that the SE system combined with a better speech assessment model can achieve better enhancement performance.

\graphicspath{ {./images/} }
\begin{figure}[t]
\centering
\includegraphics[width=7.5cm]{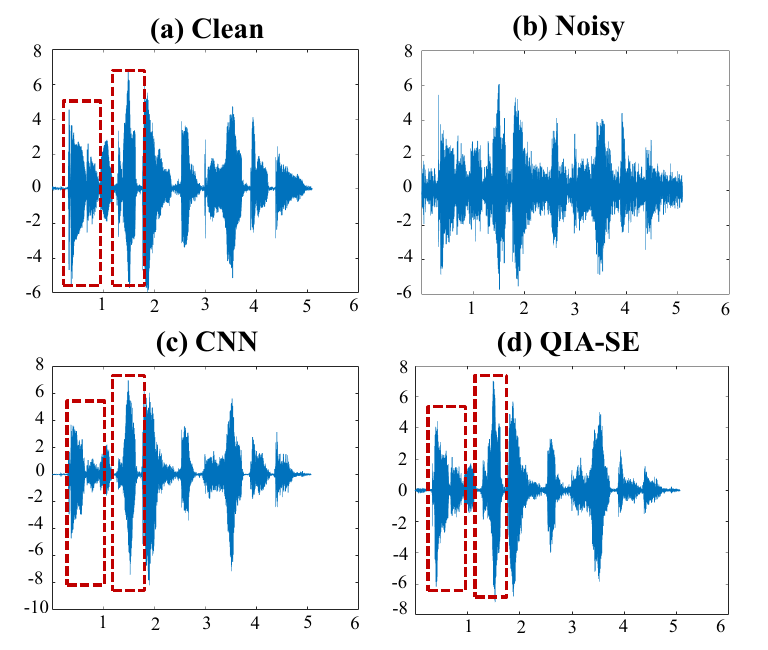} 
\caption{Waveforms of a clean utterance (Clean), its noisy version (car noise at 5 dB SNR) (Noisy), and the CNN and QIA-SE enhanced ones.} 
\label{fig:Waveform}
\end{figure}

\graphicspath{ {./images/} }
\begin{figure}[t]
\centering
\includegraphics[width=7.5cm]{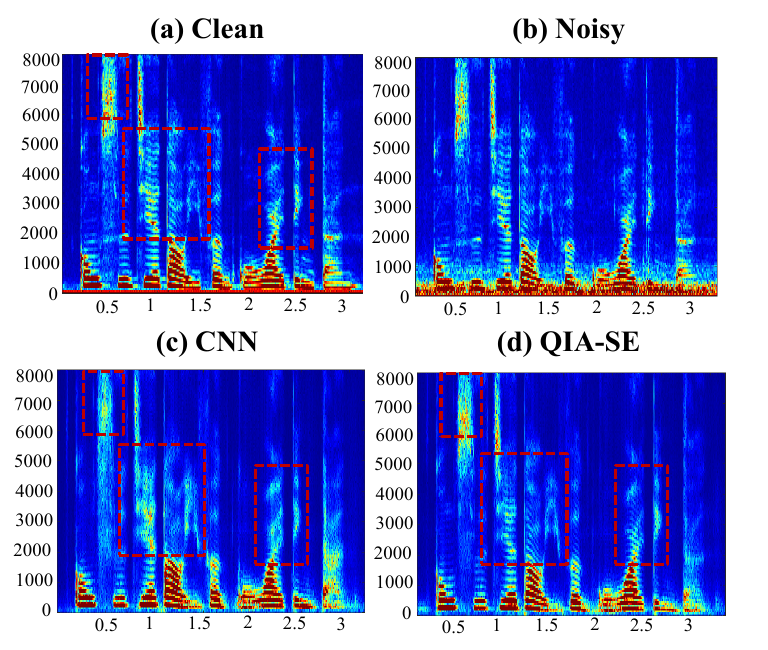} 
\caption{Spectrograms of a clean utterance (Clean), its noisy version (car noise at 5 dB SNR) (Noisy), and the CNN and QIA-SE enhanced ones.} 
\label{fig:Spectro}
\end{figure}

\graphicspath{ {./images/} }
\begin{figure}[t]
\centering
\includegraphics[width=7.5cm]{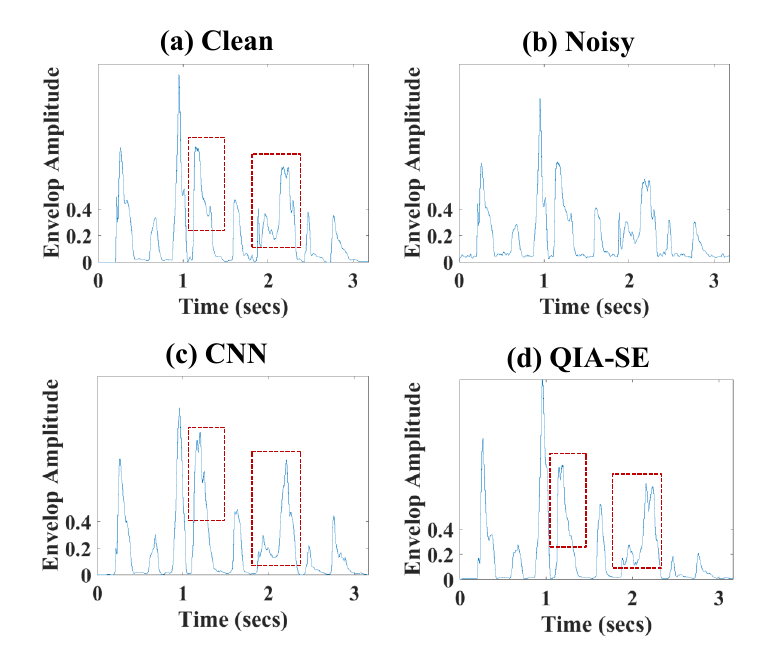} 
\caption{Amplitude envelopes from the second-channel frequency band of a clean utterance (Clean), its noisy version (car noise at 5 dB SNR) (Noisy), and the CNN and QIA-SE enhanced ones.} 
\label{fig:Envelop}
\end{figure}
\subsubsection{Qualitative Analysis}
In addition to objective evaluation, Figs. 11, 12, and 13 show the waveform, spectrogram, and amplitude envelope plots of a clean utterance, its noisy version (car noise), and the enhanced utterances (by the CNN and QIA-SE models). As shown in Fig. 11, both CNN and QIA-SE effectively removed the noise components from noisy speech. Compared with the CNN baseline, the QIA-SE model preserved more detailed structures (cf. the red rectangles in Fig. 11 (a), (c), and (d)). From Fig. 12, we also note that both CNN and QIA-SE effectively reduced noise components, and the QIA-SE model preserved more details in the spectrogram than the CNN baseline (cf. the red rectangles in Fig. 12 (a), (c), and (d)).

Several previous studies have shown that the amplitude envelope of the middle-frequency bands has a significant impact on speech intelligibility \cite{envelop, vocoder}. In this study, we adopted the four-channel tone-vocoder used in \cite{vocoder} to extract the amplitude envelope containing 457–1202 Hz components from the speech waveform. Fig. 13 shows the amplitude envelopes of the clean, noisy, and enhanced utterances processed by the CNN and QIA-SE models, where the x- and y-axes denote the time index and amplitude magnitude, respectively. The results in Fig. 13 (a), (c), and (d) show that compared with CNN enhanced speech, the amplitude envelope of QA-SE enhanced speech is more similar to that of the original clean waveform. The results further confirmed the benefits of the QIA-SE approach.

\section{CONCLUSION}
\label{sec:conclusion}
In this paper, we proposed a cross-domain speech assessment metric called MOSA-Net. We first systematically investigated the performance of the MOSA-Net with different model architectures and compared the prediction capability based on different training criteria (single-task vs multi-task training). Experimental results showed that the CRNN with the attention mechanism achieved the best performance as compared with the other models in terms of the LCC, SRCC, and MSE metrics. Next, the MOSA-Net with multi-task training consistently and significantly outperformed that with single-task training. Subsequently, we tested the MOSA-Net based on different acoustic features, including spectral features, waveform processed by learnable filter banks, and representations from SSL models. The results showed that the MOSA-Net that used cross-domain features (combining information from spectral features, complex features, raw-waveform, and SSL features) achieved the best performance. Finally, we confirmed that MOSA-Net can be used as a pre-trained model to be adapted to an assessment model for predicting subjective quality and intelligibility scores.
\par
In the second part, we proposed QIA-SE, an SE system that incorporates the information from the MOSA-Net. Experimental results showed that the QIA-SE, which directly combined the latent representations from the MOSA-Net, yielded better performance than previous SSEMS and ZMOS models, which utilized speech assessment models for offline ensemble model preparation and online model selection. In addition to better performance, the QIA-SE model required less model storage requirements and online computation. Finally, we observed that when combined with a better speech assessment model, the SE system yielded better performance. 
\par
Recently, we have used the same model architecture as MOSA-Net to predict speech intelligibility for hearing loss \cite{zezario2022mbi} and word error rate for ASR \cite{zezario2022mti}. In the future, we will explore applying the model to predict other speech assessment metrics. Moreover, how to automatically optimize the scaling factors of different losses according to specific speech processing tasks will be further investigated.
Meanwhile, we will also investigate improving the robustness of MOSA-Net in real-world scenarios. 
\bibliographystyle{IEEEbib}
\bibliography{refs}

\begin{thebibliography}{10}

\bibitem{sdi}
J.~Chen, J.~Benesty, Y.~Huang, and S.~Doclo,
\newblock ``New insights into the noise reduction {Wiener} filter,''
\newblock {\em IEEE Transactions on Audio, Speech, and Language Processing}, vol. 14, no. 4, pp. 1218--1234, 2006.

\bibitem{snr}
P.~Scalart and J.~Vieira Filho,
\newblock ``Speech enhancement based on a priori signal to noise estimation,''
\newblock in {\em Proc. ICASSP}, 1996, vol.~2, pp. 629--632.

\bibitem{ssnr}
J.~Hansen and B.~Pellom,
\newblock ``An effective quality evaluation protocol for speech enhancement algorithms,''
\newblock in {\em Proc. ICSLP}, 1998, vol.~7, pp. 2819--2822.

\bibitem{sisnr}
J.~L. Roux, S.~Wisdom, H.~Erdogan, and J.~R. Hershey,
\newblock ``{SDR} – half-baked or well done?,''
\newblock in {\em Proc. ICASSP}, 2019, pp. 626--630.

\bibitem{osisnr}
C.~Ma, D.~Li, and X.~Jia,
\newblock ``Optimal scale-invariant signal-to-noise ratio and curriculum learning for monaural multi-speaker speech separation in noisy environment,''
\newblock in {\em Proc. APSIPA ASC}, 2020, pp. 711--715.

\bibitem{ref_19}
A.~Rix, J.~Beerends, M.~Hollier, and A.~Hekstra,
\newblock ``Perceptual evaluation of speech quality ({PESQ}), an objective method for end-to-end speech quality assessment of narrow-band telephone networks and speech codecs,''
\newblock in {\em ITU-T Recommendation}, 2001, p. 862.

\bibitem{ref_20}
T.~Murphy, D.~Picovici, and A.~E. Mahdi,
\newblock ``A new single-ended measure for assessment of speech quality,''
\newblock in {\em Proc. INTERSPEECH}, 2006, pp. 177--180.

\bibitem{ref_22}
D.~Sharma, L.~Meredith, J.~Lainez, D.~Barreda, and P.~A. Naylor,
\newblock ``A non-intrusive {PESQ} measure,''
\newblock in {\em Proc. GlobalSIP}, 2014, pp. 975--978.

\bibitem{ref_25}
V.~Grancharov, D.~Y. Zhao, J.~Lindblom, and W.~B. Kleijn,
\newblock ``Low-complexity, non-intrusive speech quality assessment,''
\newblock {\em IEEE Transactions on Audio, Speech, and Language Processing}, vol. 14, pp. 1948--1956, 2006.

\bibitem{ref_26}
Q.~Li, Y.~Fang, W.~Lin, and D.~Thalmann,
\newblock ``Non-intrusive quality assessment for enhanced speech signals based on spectro temporal features,''
\newblock in {\em Proc. ICMEW}, 2014, pp. 1--6.

\bibitem{ref_32}
Q.~Li, W.~Lin, Y.~Fang, and D.~Thalmann,
\newblock ``Bag-of-words representation for non-intrusive speech quality assessment,''
\newblock in {\em Proc. ChinaSIP}, 2015, pp. 616--619.

\bibitem{ref_33}
L.~Ding, Z.~Lin, A.~Radwan, M.~S. El-Hennawey, and R.~A. Goubran,
\newblock ``Non-intrusive single-ended speech quality assessment in {VoIP},''
\newblock {\em Speech communication}, vol. 49, pp. 477--489, 2007.

\bibitem{ref_34}
F.~Rahdari, R.~Mousavi, and M.~Eftekhari,
\newblock ``An ensemble learning model for single-ended speech quality assessment using multiple-level signal decomposition method,''
\newblock in {\em Proc. ICCKE}, 2014, pp. 189--193.

\bibitem{ref_21}
T.~H. Falk and W.-Y. Chan,
\newblock ``Single-ended speech quality measurement using machine learning methods,''
\newblock {\em IEEE Transactions on Audio, Speech, and Language Processing}, vol. 14, pp. 1935--1947, 2006.

\bibitem{ref_28}
M.~Narwaria, W.~Lin, I.~V. McLoughlin, S.~Emmanuel, and L.-T. Chia,
\newblock ``Non-intrusive quality assessment of noise suppressed speech with mel-filtered energies and support vector regression,''
\newblock {\em IEEE Transactions on Audio, Speech, and Language Processing}, vol. 20, pp. 1217--1232, 2012.

\bibitem{ref_29}
M.~Narwaria, W.~Lin, I.~V. McLoughlin, S.~Emmanuel, and C.~L. Tien,
\newblock ``Non-intrusive speech quality assessment with support vector regression,''
\newblock in {\em Proc. MMM}, 2010, pp. 325--335.

\bibitem{ref_30}
T.~H. Falk, H.~Yua, and W.-Y. Chan,
\newblock ``Single-ended quality measurement of noise suppressed speech based on {Kullback Leibler} distances,''
\newblock {\em Journal of Multimedia}, vol. 2, 2007.

\bibitem{ref_31}
R.~K. Dubey and A.~Kumar,
\newblock ``Non-intrusive speech quality assessment using several combinations of auditory features,''
\newblock {\em International Journal of Speech Technology}, vol. 16, pp. 88--101, 2013.

\bibitem{ref_23}
T.-Y. Yan, M.~Wei, W.~Wei, and Z.-M. Xu,
\newblock ``A new neural network measure for objective speech quality evaluation,''
\newblock in {\em Proc. WiCOM}, 2010, pp. 1--4.

\bibitem{ref_24}
M.~Hakami and W.~B. Kleijn,
\newblock ``Machine learning based non-intrusive quality estimation with an augmented feature set,''
\newblock in {\em Proc. ICASSP}, 2017, pp. 5105--5109.

\bibitem{ref_27}
M.~H. Soni and H.~A. Patil,
\newblock ``Effectiveness of ideal ratio mask for non-intrusive quality assessment of noise suppressed speech,''
\newblock in {\em Proc. EUSIPCO}, 2017, pp. 573--577.

\bibitem{ref_35}
N.~R. French and J.~C. Steinberg,
\newblock ``Factors governing the intelligibility of speech sounds,''
\newblock {\em Journal of the Acoustical Society of America}, vol. 19, no. 1, pp. 90--119, 1947.

\bibitem{ref_36}
ANSI Std.~S3.5 1997,
\newblock ``Methods for calculation of the speech intelligibility index,''
\newblock in {\em Acoustical Society of America}, 1997.

\bibitem{ref_37}
T.~Houtgast and H.~1.~M. Steeneken,
\newblock ``Evaluation of speech transmission channels by using artificial signals,''
\newblock {\em Acustica}, vol. 25, no. 6, pp. 355--367, 1971.

\bibitem{ref_38}
H.~J.~M. Steeneken and T.~Houtgast,
\newblock ``A physical method for measuring speech-transmission quality,''
\newblock {\em Journal of the Acoustical Society of America}, vol. 67, no. 1, pp. 318--326, 1980.

\bibitem{ncm}
R.~Goldsworthy and J.~Greenberg,
\newblock ``Analysis of speech-based speech transmission index methods with implications for nonlinear operations,''
\newblock {\em Journal of the Acoustical Society of America}, vol. 116, pp. 3679--3689, 2004.

\bibitem{csii}
J.~M. Kates and K.~H. Arehart,
\newblock ``Coherence and the speech intelligibility index,''
\newblock {\em Journal of the Acoustical Society of America}, vol. 117, no. 4, pp. 2224--2237, 2005.

\bibitem{ref_39}
C.~H. Taal, R.~C. Hendriks, R.~Heusdens, and J.~Jensen,
\newblock ``An algorithm for intelligibility prediction of time-frequency weighted noisy speech,''
\newblock {\em IEEE/ACM Transactions on Audio, Speech and Language Processing}, vol. 19, no. 7, pp. 2125--2136, 2011.

\bibitem{estoi}
J.~Jensen and C.~H. Taal,
\newblock ``An algorithm for predicting the intelligibility of speech masked by modulated noise maskers,''
\newblock {\em IEEE/ACM Transactions on Audio, Speech, and Language Processing}, vol. 24, no. 11, pp. 2009--2022, 2016.

\bibitem{moda}
F.~Chen, O.~Hazrati, and P.~C. Loizou,
\newblock ``Predicting the intelligibility of reverberant speech for cochlear implant listeners with a non-intrusive intelligibility measure,''
\newblock {\em Biomedical Signal Processing and Control}, vol. 8, no. 3, pp. 311--314, 2012.

\bibitem{srmr}
T.~H. Falk, C.~Zheng, and W.~Chan,
\newblock ``A non-intrusive quality and intelligibility measure of reverberant and dereverberated speech,''
\newblock {\em IEEE Transactions on Audio, Speech, and Language Processing}, vol. 18, no. 7, pp. 1766--1774, 2010.

\bibitem{polqa_2013}
J.G Beerends, C.~Schmidmer, J.~Berger, M.~Obermann, R.~Ullmann, J.~Pomy, and M.~Keyhl,
\newblock ``Perceptual objective listening quality assessment (polqa), the third generation itu-t standard for end-to-end speech quality measurement part i—temporal alignment,''
\newblock {\em Journal of The Audio Engineering Society}, vol. 61, no. 6, pp. 366--384, june 2013.

\bibitem{somr}
N.~Mamun, M.~S.~A. Zilany, J.~H.~L. Hansen, and E.E Davies-Venn,
\newblock ``An intrusive method for estimating speech intelligibility from noisy and distorted signals,''
\newblock {\em The Journal of the Acoustical Society of America}, vol. 150, no. 3, pp. 1762--1778, 2021.

\bibitem{nopm}
N.~Mamun, W.A. Jassim, and M.~S.~A. Zilany,
\newblock ``Prediction of speech intelligibility using a neurogram orthogonal polynomial measure (nopm),''
\newblock {\em IEEE/ACM Transactions on Audio, Speech, and Language Processing}, vol. 23, no. 4, pp. 760--773, 2015.

\bibitem{nsim}
A.~Hines and N.~Harte,
\newblock ``Speech intelligibility prediction using a neurogram similarity index measure,''
\newblock {\em Speech Communication}, vol. 54, no. 2, pp. 306–320, feb 2012.

\bibitem{edraki2020speech}
A.~Edraki, W.-Y. Chan, J.~Jensen, and D.~Fogerty,
\newblock ``Speech intelligibility prediction using spectro-temporal modulation analysis,''
\newblock {\em IEEE/ACM Transactions on Audio, Speech, and Language Processing}, vol. 29, pp. 210--225, 2020.

\bibitem{rev_int}
Y.~Feng and F.~Chen,
\newblock ``Nonintrusive objective measurement of speech intelligibility: A review of methodology,''
\newblock {\em Biomedical Signal Processing and Control}, vol. 71:103204, 2022.

\bibitem{ref_43}
C.~H. Taal, R.~C. Hendriks, R.~Heusdens, and J.~Jensen,
\newblock ``An algorithm for intelligibility prediction of time-frequency weighted noisy speech,''
\newblock {\em IEEE/ACM Transactions on Audio, Speech and Language Processing}, vol. 19, no. 7, pp. 2125--2136, 2011.

\bibitem{ref_44}
J.~Ooster, R.~Huber, and B.~Meyer,
\newblock ``Prediction of perceived speech quality using deep machine listening,''
\newblock in {\em Proc. INTERSPEECH}, 2018, pp. 976--980.

\bibitem{ref_45}
P.~Seetharaman, G.~Mysore, P.~Smaragdis, and B.~Pardo,
\newblock ``Blind estimation of the speech transmission index for speech quality prediction,''
\newblock in {\em Proc. ICASSP}, 2018, pp. 591--595.

\bibitem{ref_46}
J.~Ooster and B.~Meyer,
\newblock ``Improving deep models of speech quality prediction through voice activity detection and entropy based measures,''
\newblock in {\em Proc. ICASSP}, 2019, pp. 636--640.

\bibitem{ref_47}
H.~Gamper, C.~Reddy, R.~Cutler, I.~J. Tashev, and J.~Gehrke,
\newblock ``Intrusive and nonintrusive perceptual speech quality assessment using a convolutional neural network,''
\newblock in {\em Proc. WASPAA}, 2019, pp. 85--89.

\bibitem{ref_48}
A.~R. Avila, H.~Gamper, C.~Reddy, R.~Cutler, I.~Tashev, and J.~Gehrke,
\newblock ``Non-intrusive speech quality assessment using neural networks,''
\newblock in {\em Proc. ICASSP}, 2019, pp. 631--635.

\bibitem{dnsmos}
C.~K.~A. Reddy, V.~Gopal, and R.~Cutler,
\newblock ``{DNSMOS}: A non-intrusive perceptual objective speech quality metric to evaluate noise suppressors,''
\newblock in {\em Proc. ICASSP}, 2021, pp. 6493--6497.

\bibitem{ref_49}
S.-W. Fu, Y.~Tsao, H.-T. Hwang, and H.-W. Wang,
\newblock ``{Quality-Net}: An end-to-end non-intrusive speech quality assessment model based on {BLSTM},''
\newblock in {\em Proc. INTERSPEECH}, 2018, pp. 1873--1877.

\bibitem{ref_55}
X.~Jia and D.~Li,
\newblock ``A deep learning-based time-domain approach for non-intrusive speech quality assessment,''
\newblock in {\em Proc. APSIPA ASC}, 2020, pp. 477--481.

\bibitem{ref_56}
X.~Dong and D.~S. Williamson,
\newblock ``An attention enhanced multi-task model for objective speech assessment in real-world environments,''
\newblock in {\em Proc. ICASSP}, 2020, pp. 911--915.

\bibitem{ref_52}
R.~E. Zezario, S.-W. Fu, C.-S Fuh, Y.~Tsao, and H.-M. Wang,
\newblock ``{STOI-Net}: A deep learning based non-intrusive speech intelligibility assessment model,''
\newblock in {\em Proc. APSIPA ASC}, 2020, pp. 482--486.

\bibitem{mbnet_mos}
Y.~Leng, X.~Tan, S.~Zhao, F.~Soong, X.-Y. Li, and T.~Qin,
\newblock ``{MBNet}: {MOS} prediction for synthesized speech with mean-bias network,''
\newblock in {\em Proc. ICASSP}, 2021, pp. 391--395.

\bibitem{choi2020neural}
Y.~Choi, Y.~Jung, and H.~Kim,
\newblock ``Neural {MOS} prediction for synthesized speech using multi-task learning with spoofing detection and spoofing type classification,''
\newblock in {\em Proc. SLT}, 2020, pp. 462--469.

\bibitem{hu2021svsnet}
C.~H. Hu, Y.-H. Peng, J.~Yamagishi, Y.~Tsao, and H.-M. Wang,
\newblock ``{SVSNet}: An end-to-end speaker voice similarity assessment model,''
\newblock {\em IEEE Signal Processing Letters}, vol. 29, pp. 767--771, 2022.

\bibitem{wav2vec_mos}
W.-C. Tseng, C.~y.~Huang, W.-T. Kao, Y.~Lin, and H.~y.~Lee,
\newblock ``Utilizing self-supervised representations for {MOS} prediction,''
\newblock in {\em Proc. INTERSPEECH}, 2021, pp. 2781--2785.

\bibitem{sincnet}
M.~Ravanelli and Y.~Bengio,
\newblock ``Speaker recognition from raw waveform with {SincNet},''
\newblock in {\em Proc. SLT}, 2018.

\bibitem{wav2vec}
A.~Baevski, Y.~Zhou, A.~Mohamed, and M.~Auli,
\newblock ``{Wav2vec 2.0}: A framework for self-supervised learning of speech representations,''
\newblock in {\em Proc. NIPS}, 2020.

\bibitem{hubert}
W.-N. Hsu, B.~Bolte, Y.-Hung~Hubert Tsai, K.~Lakhotia, R.~Salakhutdinov, and A.~Mohamed,
\newblock ``{HuBERT}: Self-supervised speech representation learning by masked prediction of hidden units,''
\newblock {\em IEEE/ACM Transactions on Audio, Speech, and Language Processing}, 2021.

\bibitem{ssl-mos}
E.~Cooper, W.-H. Huang, T.~Toda, and J.~Yamagishi,
\newblock ``Generalization ability of mos prediction networks,''
\newblock in {\em Proc. ICASSP}, 2022.

\bibitem{fu2019metricGAN}
S.-W. Fu, C.-F. Liao, Y.~Tsao, and S.-D. Lin,
\newblock ``{MetricGAN}: Generative adversarial networks based black-box metric scores optimization for speech enhancement,''
\newblock in {\em Proc. ICML}, 2019, vol.~97, pp. 2031--2041.

\bibitem{metricgan+}
S.-W. Fu, C.~Yu, T.-A. Hsieh, and et.al.,
\newblock ``{MetricGAN+}: An improved version of {MetricGAN} for speech enhancement,''
\newblock in {\em Proc. INTERSPEECH}, 2021, pp. 201--205.

\bibitem{DeepNoise}
Z.~Xu, M.~Strake, and T.~Fingscheidt,
\newblock ``Deep noise suppression maximizing non-differentiable pesq mediated by a non-intrusive pesqnet,''
\newblock {\em IEEE/ACM Transactions on Audio, Speech, and Language Processing}, vol. 30, pp. 1572--1585, 2022.

\bibitem{metricganu}
S.-W. Fu, C.~Yu, K.-H. Hung, M.~Ravanelli, and Y.~Tsao,
\newblock ``Metricgan-u: Unsupervised speech enhancement/ dereverberation based only on noisy/ reverberated speech,''
\newblock in {\em Proc. ICASSP}, 2022.

\bibitem{zezario2019specialized}
R.~E. Zezario, S.-W. Fu, X.~Lu, H.-M. Wang, and Y.~Tsao,
\newblock ``Specialized speech enhancement model selection based on learned non-intrusive quality assessment metric,''
\newblock in {\em Proc. INTERSPEECH}, 2019, pp. 3168--3172.

\bibitem{ref_54}
R.~E. Zezario, C.-S. Fuh, H.-M. Wang, and Y.~Tsao,
\newblock ``Speech enhancement with zero-shot model selection,''
\newblock in {\em Proc. EUSIPCO}, 2021.

\bibitem{ref_51}
C.-C. Lo, S.-W. Fu, W.-C. Huang, X.~Wang, J.~Yamagishi, Y.~Tsao, and H.-M. Wang,
\newblock ``{MOSNet}: deep learning-based objective assessment for voice conversion,''
\newblock in {\em Proc. INTERSPEECH}, 2019, pp. 1541--1545.

\bibitem{andersen2018nonintrusive}
A.~H. Andersen, J.~M.~De Haan, Z.~H Tan, and J.~Jensen,
\newblock ``Nonintrusive speech intelligibility prediction using convolutional neural networks,''
\newblock {\em IEEE/ACM Transactions on Audio, Speech, and Language Processing}, vol. 26, no. 10, pp. 1925--1939, 2018.

\bibitem{SIP_DL}
M.~B. Pedersen, A.~Heidemann Andersen, S.~H. Jensen, and J.~Jensen,
\newblock ``A neural network for monaural intrusive speech intelligibility prediction,''
\newblock in {\em Proc. ICASSP}, 2020, pp. 336--340.

\bibitem{MONO99}
G.~A. Studebaker, R.~L. Sherbecoe, D.~M. McDaniel, and C.~A Gwaltney,
\newblock ``Monosyllabic word recognition at higher-than-normal speech and noise levels,''
\newblock {\em The Journal of the Acoustical Society of America}, vol. 105, no. 4, pp. 2431--2444, 1999.

\bibitem{dubno2005word}
J.~R. Dubno, A.~R Horwitz, and J.~B Ahlstrom,
\newblock ``Word recognition in noise at higher-than-normal levels: Decreases in scores and increases in masking,''
\newblock {\em The Journal of the Acoustical Society of America}, vol. 118, no. 2, pp. 914--922, 2005.

\bibitem{POLQA}
G.~Mittag and S.~Möller,
\newblock ``Non-intrusive speech quality assessment for super-wideband speech communication networks,''
\newblock in {\em Proc. ICASSP}, 2019, pp. 7125--7129.

\bibitem{katehaspi}
J.~M. Kates and K.~H. Arehart,
\newblock ``The hearing-aid speech perception index ({HASPI}) version 2,''
\newblock {\em Speech Communication}, vol. 131, pp. 35--46, 2021.

\bibitem{pyramidlstm}
X.~Dong and D.~S. Williamson,
\newblock ``A pyramid recurrent network for predicting crowdsourced speech-quality ratings of real-world signals,''
\newblock in {\em Proc. INTERSPEECH}, 2020, pp. 4631--4635.

\bibitem{end2endmultisubject}
Z.~Zhang, P.~Vyas, X.~Dong, and D.~S. Williamson,
\newblock ``An end-to-end non-intrusive model for subjective and objective real-world speech assessment using a multi-task framework,''
\newblock in {\em Proc. ICASSP}, 2021, pp. 316--320.

\bibitem{ref_58}
F.-K. Chuang, S.-S. Wang, J.~w.~Hung, Y.~Tsao, and S.-H. Fang,
\newblock ``Speaker-aware deep denoising autoencoder with embedded speaker identity for speech enhancement,''
\newblock in {\em Proc. INTERSPEECH}, 2019, pp. 3173--3177.

\bibitem{ref_59}
Y.~Koizumi, K.~Yatabe, M.~Delcroix, Y.~Masuyama, and D.~Takeuchi,
\newblock ``Speech enhancement using self-adaptation and multi-head self-attention,''
\newblock in {\em Proc. ICASSP}, 2020, pp. 181--185.

\bibitem{ref_60}
M.~Delcroix, K.~Zmolikova, K.~Kinoshita, A.~Ogawa, and T.~Nakatani,
\newblock ``Single channel target speaker extraction and recognition with speaker beam,''
\newblock in {\em Proc. ICASSP}, 2018, pp. 5554--5558.

\bibitem{ref_61}
K.~Zmolikova, M.~Delcroix, K.~Kinoshita, T.~Ochiai, T.~Nakatani, L.~Burget, and J.~Cernocky,
\newblock ``{SpeakerBeam}: Speaker aware neural network for target speaker extraction in speech mixtures,''
\newblock {\em IEEE Journal of Selected Topics in Signal Processing}, vol. 13, no. 4, pp. 800--814, 2019.

\bibitem{ref_62}
J.~Lee, Y.~Jung, M.~Jung, and H.~Kim,
\newblock ``Dynamic noise embedding: Noise aware training and adaptation for speech enhancement,''
\newblock {\em arXiv:2008.11920}, 2020.

\bibitem{nayem21_interspeech}
K.~M. Nayem and D.~S. Williamson,
\newblock ``Incorporating embedding vectors from a human mean-opinion score prediction model for monaural speech enhancement,''
\newblock in {\em Proc. INTERSPEECH}, 2021, pp. 216--220.

\bibitem{chang2021moevc}
Y.-T. Chang, Y.~H. Yang, Y.-H. Peng, S.-S.~Wang ang, T.-S. Chi, Y.~Tsao, and H.~M. Wang,
\newblock ``{MoEVC}: A mixture of experts voice conversion system with sparse gating mechanism for online computation acceleration,''
\newblock in {\em Proc. ISCSLP}, 2021, pp. 1--5.

\bibitem{ref_63}
D.~Paul and J.~Baker,
\newblock ``The design for the {Wall Street Journal}-based {CSR} corpus,''
\newblock in {\em Proc. ICSLP}, 1992, pp. 899--902.

\bibitem{ref_64}
D.~Hu and D.~Wang,
\newblock ``Pnl 100 nonspeech sounds[online],'' \url{http://web.cse.ohio-state.edu/pnl/corpus/HuNonspeech/HuCorpus.html}, 2010.

\bibitem{srcc}
C.~Spearman,
\newblock ``The proof and measurement of association between two things,''
\newblock {\em The American Journal of Psychology}, vol. 15, no. 1, pp. 72--101, 1904.

\bibitem{KingmaB14}
D.P. Kingma and J.Ba,
\newblock ``Adam: A method for stochastic optimization,''
\newblock in {\em Proc. ICLR}, 2015.

\bibitem{tieleman2012lecture}
T.~Tieleman, G.~Hinton, et~al.,
\newblock ``Lecture 6.5-rmsprop: Divide the gradient by a running average of its recent magnitude,''
\newblock {\em COURSERA: Neural networks for machine learning}, vol. 4, no. 2, pp. 26--31, 2012.

\bibitem{TIMIT}
J.~Garofolo, L.~Lamel, W~Fisher, J.~Fiscus, D.~Pallett, and N.~Dahlgren,
\newblock ``Darpa timit acoustic-phonetic continuous speech corpus cd-rom {TIMIT},'' 1993-02-01 1993.

\bibitem{hu2007evaluation}
Yi~Hu and Philipos~C Loizou,
\newblock ``Evaluation of objective quality measures for speech enhancement,''
\newblock {\em IEEE Transactions on audio, speech, and language processing}, vol. 16, no. 1, pp. 229--238, 2007.

\bibitem{TMINT-QI}
Y.-W. Chen and Y.~Tsao,
\newblock ``{InQSS}: a speech intelligibility assessment model using a multi-task learning network,''
\newblock in {\em To appear in Proc. INTERSPEECH}, 2022.

\bibitem{klt}
A.~Rezayee and S.~Gazor,
\newblock ``An adaptive {KLT} approach for speech enhancement,''
\newblock {\em IEEE Transactions on Speech and Audio Processing}, vol. 9, no. 2, pp. 87--95, 2001.

\bibitem{mmse}
Y.~Ephraim and D.~Malah,
\newblock ``Speech enhancement using a minimum mean-square error log-spectral amplitude estimator,''
\newblock {\em IEEE Transactions on Acoustics, Speech, and Signal Processing}, vol. 33, no. 2, pp. 443--445, 1985.

\bibitem{FCN}
S.-W. Fu, Y.~Tsao, X.~Lu, and H.~Kawai,
\newblock ``Raw waveform-based speech enhancement by fully convolutional networks,''
\newblock in {\em Proc. APSIPA ASC}, 2017.

\bibitem{DDAE}
X.~Lu, Y.~Tsao, S.~Matsuda, and C.~Hori,
\newblock ``Speech enhancement based on deep denoising autoencoder,''
\newblock in {\em Proc. INTERSPEECH}, 2013, pp. 436--440.

\bibitem{Trans}
J.~Kim, M.~El-Khamy, and J.~Lee,
\newblock ``{T-GSA}: Transformer with {Gaussian}-weighted self-attention for speech enhancement,''
\newblock in {\em Proc. ICASSP}, 2020, pp. 6649--6653.

\bibitem{ref_67}
M.~Huang,
\newblock ``Development of {Taiwan} {Mandarin} hearing in noise test,''
\newblock {\em Department of speech language pathology and audiology, National Taipei University of Nursing and Health Science}, 2005.

\bibitem{envelop}
R.~van Hoesel, M.~Böhm, R.D. Battmer, J.~Beckschebe, and T.~Lenarz,
\newblock ``Amplitude-mapping effects on speech intelligibility with unilateral and bilateral cochlear implants,''
\newblock vol. 24, no. 4, pp. 381--382, 2005.

\bibitem{vocoder}
S.-W. Fu, P.-C. Li, Y.-H. Lai, C.-C. Yang, L.-C. Hsieh, and Y.~Tsao,
\newblock ``Joint dictionary learning-based non-negative matrix factorization for voice conversion to improve speech intelligibility after oral surgery,''
\newblock {\em IEEE Transactions on Biomedical Engineering}, vol. 64, no. 11, pp. 2584--2594, 2017.

\bibitem{zezario2022mbi}
R.~E. Zezario, F.~Chen, C.~S. Fuh, H.-M. Wang, and Y.~Tsao,
\newblock ``Mbi-net: A non-intrusive multi-branched speech intelligibility prediction model for hearing aids,''
\newblock in {\em To appear in Proc. INTERSPEECH}, 2022.

\bibitem{zezario2022mti}
R.~Zezario, S.-W. Fu, F.~Chen, C.~S. Fuh, H.-M. Wang, and Y.~Tsao,
\newblock ``Mti-net: A multi-target speech intelligibility prediction model,''
\newblock in {\em To appear in Proc. INTERSPEECH}, 2022.

\end{thebibliography}
\end{document}